\documentclass[journal,review,submit,pdftex,moreauthors]{Definitions/mdpi}

\firstpage{1}
\pubvolume{1}
\issuenum{1}
\articlenumber{0}
\pubyear{2024}
\copyrightyear{2024}
\datereceived{ }
\daterevised{ } 
\dateaccepted{ }
\datepublished{ }
\usepackage{subcaption}
\makeatletter
\renewcommand{\linenumbers}{}
\makeatother

\hreflink{https://doi.org/} 

\Title{The Observed Luminosity Correlations of Gamma-Ray Bursts and Their Applications}

\TitleCitation{GRB correlations}

\Author{Chen Deng $^{1}$ \orcidA, Yong-Feng Huang $^{1,2,}$*
\orcidB, Fan Xu $^{3}$ \orcidC and Abdusattar Kurban$^{4,5,6}$
\orcidD}

\AuthorNames{Chen Deng, Yong-Feng Huang, Fan Xu and Abdusattar
Kurban}

\AuthorCitation{Deng C.; Huang Y.-F.; Xu F.; Kurban A.}

\address{%
$^{1}$ \quad School of Astronomy and Space Science, Nanjing
University, Nanjing 210023, China \\
$^{2}$ \quad Key Laboratory of Modern Astronomy and Astrophysics
(Nanjing University), Ministry of Education, China \\
$^{3}$ \quad Department of Physics, Anhui Normal University, Wuhu 241002, China \\
$^{4}$ \quad Xinjiang Astronomical Observatory, Chinese Academy
of Sciences, Urumqi 830011, Xinjiang, China \\
$^{5}$ \quad Key Laboratory of Radio Astronomy and Technology
(Chinese Academy of Sciences), Beijing 100101, China \\
$^{6}$ \quad Xinjiang Key Laboratory of Radio Astrophysics, Urumqi
830011, Xinjiang, China}

\corres{Correspondence: hyf@nju.edu.cn}

\abstract{Gamma-ray bursts (GRBs) are among the most luminous
electromagnetic transients in the universe, providing unique
insights into extreme astrophysical processes and serving as
promising probes for cosmology. Unlike Type Ia supernovae, which
have a unified explosion mechanism, GRBs cannot directly act as
standard candles for tracing cosmic evolution at high redshifts
due to significant uncertainties in their underlying physical
origins. Empirical correlations derived from statistical analyses
involving various GRB parameters provide valuable information
regarding their intrinsic properties. In this brief review, we
describe various correlations among GRB parameters involving the
prompt and afterglow phases, discussing possible theoretical
interpretations behind them. The scarcity of low-redshift
GRBs poses a major obstacle to the application of GRB empirical
correlations in cosmology, referred to as the circularity problem.
We present various efforts aiming at calibrating GRBs to address
this challenge and leveraging established empirical correlations
to constrain cosmological parameters. The pivotal role of GRB
sample quality in advancing cosmological research is underscored.
Some correlations that could potentially be utilized as redshift
indicators are also introduced.}


\begin{document}


\section{Introduction}

Gamma-ray bursts (GRBs) are one of the most energetic stellar
explosion events in the universe, with the isotropic equivalent
energies ranging from $10^{48}$ -- $10^{55}$ erg
\cite{Kumar:2015PhR...561....1K,
Gehrels:2013FrPhy...8..661G,Nakar:2007PhR...442..166N,Zhang:2018pgrb.book.....Z,
Berger:2014ARA&A..52...43B,Piran:2004RvMP...76.1143P}. Since their
discovery in the 1960s, the use of various observational
instruments has led to groundbreaking advancements in the field of
GRBs, significantly deepening our understanding of their physical
nature
\cite{Boella:1997A&AS..122..299B,Fishman:1994ApJS...92..229F,
Aptekar:1995SSRv...71..265A,Ubertini:2003A&A...411L.131U,Gehrels:2004ApJ...611.1005G,
Barthelmy:2005SSRv..120..143B,Meegan:2009ApJ...702..791M}. The
Burst and Transient Source Experiment (BATSE) on board the Compton
Gamma Ray Observatory (CGRO) revealed an isotropic spatial
distribution of GRBs, suggesting their cosmological origin
\cite{Briggs:1996ApJ...459...40B,
Balazs:1998A&A...339....1B,Tarnopolski:2017MNRAS.472.4819T,Vavrek:2008MNRAS.391.1741V,
Meszaros:2000ApJ...539...98M}. This was unequivocally confirmed by
the detection of the multi-wavelength afterglow of GRB 970508,
leading to the first redshift measurement of $z=0.835$ for it
\cite{Metzger:1997Natur.387..878M}.

Observationally, the duration of GRBs' prompt emission,
represented by $T_{90}$, is defined as the time interval during
which $5\%$--$95\%$ of gamma-ray photons are detected. This
duration exhibits a bimodal distribution, with approximately 2
seconds as the dividing line
\cite{Kouveliotou:1993ApJ...413L.101K}. Long GRBs (LGRBs,
$T_{90}>2$s) are found in star-forming galaxies and are associated
with Type Ic supernovae
\cite{Hjorth:2003Natur.423..847H,Malesani:2004ApJ...609L...5M,
Woosley:2006ARA&A..44..507W,Sparre:2011ApJ...735L..24S,Schulze:2014A&A...566A.102S,
Campana:2006Natur.442.1008C,Bloom:1999Natur.401..453B,Galama:1998Natur.395..670G,
Fruchter:2006Natur.441..463F,Prochaska:2007ApJ...666..267P,Rueda:2021ARep...65.1026R}
This indicates that LGRBs originate from the collapse of massive stars,
with Wolf-Rayet stars as the likely progenitors
\cite{Woosley:1993ApJ...405..273W,MacFadyen:1999ApJ...524..262M,Crowther:2007ARA&A..45..177C,Bromberg:2012ApJ...749..110B,Kelly:2008ApJ...687.1201K,Raskin:2008ApJ...689..358R}. On
average, short GRBs (SGRBs, $T_{90}<2$s) display a larger offset
relative to their host galaxies compared to LGRBs
\cite{Berger:2010ApJ...722.1946B,Fong:2010ApJ...708....9F,
Fong:2013ApJ...776...18F,Fong:2013ApJ...769...56F}. Additionally,
their optical/infrared afterglows may contain a kilonova
component, as predicted by the merger of neutron star - neutron
star (NS-NS) binaries or neutron star - black hole (NS-BH)
binaries \cite{Li:1998ApJ...507L..59L,Tanvir:2017ApJ...848L..27T,
Smartt:2017Natur.551...75S,Arcavi:2017Natur.551...64A,Metzger:2017LRR....20....3M,
Rastinejad:2022Natur.612..223R,Yang:2022Natur.612..232Y,Chen:2024ApJ...971..143C}.
The association of GRB 170817A with the gravitational wave event
GW170817 detected by the advanced Laser Interferometer
Gravitational Wave Observatory provids strongly evidence that
binary neutron star mergers are one of the progenitors of SGRBs
\cite{Abbott:2017PhRvL.119p1101A,Abbott:2017ApJ...848L..12A,
Abbott:2017ApJ...848L..13A}. Interestingly, some GRBs exhibit
characteristics of both long and short GRBs. For example, GRB
211211A, though classified as a LGRB, is associated with a
kilonova -- a phenomenon typically linked to SGRBs
\cite{Rastinejad:2022Natur.612..223R,Yang:2022Natur.612..232Y,Rossi:2022ApJ...932....1R}.
The classification of GRBs solely based on their duration may lead
to confusion. This is because the prompt emission duration is
energy and sensitivity dependent, meaning that the duration may
vary across different detectors
\cite{Bromberg:2013ApJ...764..179B}. The overlap in the duration
distributions of long and short GRBs suggests the possible
existence of an intermediate-duration group
\cite{Horvath:1998ApJ...508..757H,
Mukherjee:1998ApJ...508..314M,Horvath:2002A&A...392..791H,Veres:2010ApJ...725.1955V}.
Furthermore, the discovery of some subclasses, such as SGRBs with
extended emission and ultra-long GRBs, has introduced challenges
to the traditional GRB classification
\cite{Norris:2006ApJ...643..266N,Virgili:2013ApJ...778...54V,Levan:2014ApJ...781...13L,Boer:2015ApJ...800...16B,DellaValle:2006Natur.444.1050D}. To address these complexities, machine
learning methods have been applied to enhance GRB classification
\cite{Jespersen:2020ApJ...896L..20J,
Garcia-Cifuentes:2023ApJ...951....4G}. In addition, Zhang et al.
\cite{Zhang:2006Natur.444.1010Z,Kann:2011ApJ...734...96K} proposed a physical
classification system that divides GRBs into Type I and Type II by
integrating multiple criteria, including spectral information,
empirical correlations, host galaxy characteristics, and
associations with supernova or kilonova.

A GRB generally consists of two phases: the prompt emission phase
and the afterglow phase. The prompt emission phase can last from
sub-seconds to thousands of seconds
\cite{Horvath:2006A&A...447...23H}, with light curves that exhibit
highly irregular and rapid variations
\cite{Fishman:1994ApJS...92..229F}. This phase may also contain
multiple emission episodes
\cite{Zhang:2012ApJ...748..132Z,Hu:2014ApJ...789..145H,
Lazzati:2005MNRAS.357..722L}, which includes quiescent stages in
which the flux drops to nearly background level. The pulse
profiles of the prompt emission generally display an asymmetric
shape, commonly described by a ``fast-rise exponential decay”
(FRED) pattern
\cite{Norris:1996ApJ...459..393N,Norris:2005ApJ...627..324N}. The
prompt emission spectrum of GRBs can be phenomenologically
characterized by the so-called Band function
\cite{Band:1993ApJ...413..281B}, with typical photon spectral
indices of $-1$ and $-2$ for the low-energy and high-energy
segments \cite{Preece:2000ApJS..126...19P}, respectively. In
addition, some GRBs show thermal or high-energy components in
their prompt phase
\cite{Ryde:2005ApJ...625L..95R,Ryde:2009ApJ...702.1211R,
Ryde:2010ApJ...709L.172R,Chen:2024ApJ...972..132C,Abdo:2009Sci...323.1688A,
Abdo:2009ApJ...706L.138A,Ackermann:2010ApJ...716.1178A,Ackermann:2011ApJ...729..114A}.
The spectral evolution of GRBs usually follows two patterns
\cite{Golenetskii:1983Natur.306..451G,
Norris:1986ApJ...301..213N,Lu:2012ApJ...756..112L}: one is the
``hard-to-soft'' pattern, where the peak photons gradually shifts
from high to low energies, and the other is the ``flux-tracking''
pattern, in which the spectral evolution synchronizes with the
flux variations. In some GRBs, a transition between these spectral
evolution modes can occur \cite{Deng:2024ApJ...970...67D}, further
highlighting the diversity and complexity of GRB radiation
processes. Following the prompt emission, the afterglow phase
ensues, which can last for months or even longer. The
multi-wavelength afterglow emission covers the entire
electromagnetic spectrum, and the afterglow light curves are
typically smoother than those of the prompt phase, often described
by a multi-segment power-law function. Thanks to the
\textit{Swift} satellite's rapid slewing capability, some elusory
components in early X-ray afterglows have been revealed
\cite{Burrows:2005Sci...309.1833B,
Nousek:2006ApJ...642..389N,O'Brien:2006ApJ...647.1213O}. With the
accumulation of the observational data, Zhang et al.
\cite{Zhang:2006ApJ...642..354Z} summarized a canonical X-ray
light curve with five components, noting that GRBs may exhibit
some or even all of these features. Additionally, an analogous
hypothetical optical light curve with richer features has also
been proposed \cite{Li:2012ApJ...758...27L}.

The so called ``fireball'' model has been established for GRBs to
explain the rich observational phenomena in both the prompt and
afterglow phases
\cite{Rees:1992MNRAS.258P..41R,Piran:1993MNRAS.263..861P,Wijers:1997MNRAS.288L..51W}.
Following the catastrophic destruction of the progenitor star, a
hyper-accreting black hole or highly magnetized neutron star is
left, which could act as the central engine of a GRB
\cite{Dai:1998PhRvL..81.4301D,Zhang:2001ApJ...552L..35Z,Fryer:2014ApJ...793L..36F,Fryer:2015PhRvL.115w1102F}. The
central engine launches a series of ultra-relativistic shells with
different speed. These shells collide at distances of
approximately $10^{14}$ to $10^{16}$ cm from the central engine,
generating internal shocks that accelerate electrons to produce
optically thin synchrotron radiation
\cite{Rees:1994ApJ...430L..93R,Kobayashi:1997ApJ...490...92K,Daigne:1998MNRAS.296..275D,
Geng:2018ApJS..234....3G}. Additionally, the photosphere model,
another explanation for the prompt emission, has also been widely
discussed. When additional effects are considered, this model can
also account for non-thermal spectrum observed in some GRBs
\cite{Pe'er:2011ApJ...732...49P,Bhattacharya:2020MNRAS.491.4656B}.
As the merged shells sweep up the surrounding medium, external
forward shocks are formed, leading to the production of
multi-wavelength afterglows, spanning from radio waves to gamma
rays \cite{Sari:1998ApJ...497L..17S}. However, despite the wide
acceptance of the fireball model, our understanding of the
physical mechanisms underlying GRBs, particularly during the
prompt emission phase, remains limited. Synchrotron radiation and
synchrotron self-Compton scattering can account for some spectral
features of GRBs, which predict that the low-energy spectral index
should not exceed $-2/3$, a threshold known as the synchrotron
``line of death'' \cite{Preece:1998ApJ...506L..23P}. Nevertheless,
in some GRBs, the spectral index $\alpha$ does exceed $-2/3$,
suggesting that the prompt emission likely arises from a
combination of multiple, complex radiation processes
\cite{Nava:2011A&A...530A..21N,Nava:2011MNRAS.415.3153N}. The
fireball dynamics is well described by the Blandford-McKee
self-similar solution \cite{Blandford:1976PhFl...19.1130B}, which,
combined with the slow-cooling synchrotron radiation mechanism,
can effectively explain the power-law decay behavior of the
observed X-ray afterglow. However, the X-ray/internal plateaus
seen in some afterglows deviate from the predictions of the
standard forward shock model. To address this discrepancy,
researchers have proposed several models to explain the plateau
phase \cite{Dai:1998PhRvL..81.4301D,
Ioka:2006A&A...458....7I,Racusin:2008Natur.455..183R,Dereli-Begue:2022NatCo..13.5611D,
Oganesyan:2020ApJ...893...88O}.

Since GRBs are extremely luminous, they can be detected at high
redshifts. The highest redshift record of detected GRBs is 9.4
\cite{Cucchiara:2011ApJ...736....7C,Salvaterra:2009Natur.461.1258S}. So, GRBs can be potentially
used as robust cosmological probes. GRBs offer a unique
opportunity to investigate the history of star formation, metal
enrichment, and cosmic reionization at high redshift (see Ref.
\cite{Wang:2015NewAR..67....1W} and references therein). However,
unlike Type Ia supernovae (SNe Ia), which have a uniform explosion
mechanism, the physical processes behind GRBs are not fully
understood. Additionally, the isotropic energy of GRBs spans
several orders of magnitude, making it challenging to directly use
them as standard candles to explore the high-redshift universe. In
this context, identifying universal observational correlations
that can standardize GRBs becomes particularly important. These
empirical correlations offer critical insights into GRB
classification, the emission mechanisms, and the structure and
evolution of GRB jets. Here we present a review on various aspects
of GRBs, paying special attention on the correlations involving
their luminosity.

This article is organized as follows. In Section \ref{sec2}, we
provide a brief overview of the discovery and development of some
empirical correlations concerning the parameters of the prompt
phase and afterglow phase, and discuss possible physical
interpretations underlying these correlations. Section \ref{sec3}
presents the application of some empirical correlations in
constraining cosmological parameters. The usage of some
correlations as pseudo-redshift indicators is also described.
Finally, Section \ref{sec4} presents our conclusions.


\section{GRB Prompt and Afterglow Correlations}\label{sec2}
This section provides a brief summary of common empirical
correlations related to GRB prompt and afterglow parameters,
followed by a discussion of their potential physical origins. For
comprehensive reviews on GRB prompt and afterglow correlations,
the readers could refer to Refs.
\cite{Dainotti:2017NewAR..77...23D,Dainotti:2018AdAst2018E...1D,
Dainotti:2019gbcc.book.....D,Parsotan:2022Univ....8..310P}.

\subsection{GRB prompt correlations}
\subsubsection{The $L_{\rm p}-\tau_{\rm lag}$ correlation}
Norris et al. \cite{Norris:2000ApJ...534..248N} first identified
an inverse correlation between the peak luminosity ($L_{\rm p}$)
and spectral lag ($\tau_{\rm lag}$) based on a sample of six BATSE
GRBs with known redshifts, described as
\begin{eqnarray}
\frac{L_{\rm p}}{10^{53}\rm erg~s^{-1}} & = & 1.3(\frac{\tau_{\rm lag}}
{0.01~\rm s})^{-1.14},
\end{eqnarray}
where $L_{\rm p}$ denotes the isotropic equivalent peak
luminosity in 50 -- 300 keV and $\tau_{\rm lag}$ is defined as
the delay between BATSE channels 1 and 3, which correspond to the
energy ranges of 25-50 keV and 50-300 keV, respectively. This
finding is confirmed by the analysis results reported by Schaefer
et al. \cite{Schaefer:2001ApJ...563L.123S}, who calculated the
peak flux using 256 ms time bin for 50--300keV, based on 112 BATSE
GRBs with no measured redshifts. Tsutsui et al.
\cite{Tsutsui:2008MNRAS.386L..33T} investigated the $L_{\rm
p}-\tau_{\rm lag}$ correlation with pseudo-redshifts derived from
the Yonetoku relation for 565 BATSE GRBs, noting a substantial
dispersion with a correlation coefficient of only 0.38.
Furthermore, a redshift-dependent $L_{\rm p}-\tau_{\rm lag}$
correlation was proposed as
\begin{eqnarray}
\frac{L_{\rm p}}{10^{52}\rm erg~s^{-1}} = 10^{-1.12 \pm 0.07} (1+z)^{2.53 \pm 0.10}
\tau_{\text{lag}}^{-0.28 \pm 0.03}.
\end{eqnarray}

In contrast to the spectral lag extracted from different energy
channels in the observer frame, Ukwatta et al.
\cite{Ukwatta:2012MNRAS.419..614U} analyzed a sample of 43
\textit{Swift} GRBs with known redshifts to update the
luminosity-lag correlation in the GRB source frame. Their result
is characterized by a power-law index of -1.2, i.e., $L_{\rm
p}\propto (\tau_{\rm lag}/(1+z))^{-1.2 \pm 0.2}$. The energy
channels used to compute the spectral lag in their study are
100-150 keV and 200-250 keV for \textit{Swift} BAT in the GRB rest
frame. In addition, similar luminosity-lag relation was also
revealed in X-ray flares of GRB afterglows
\cite{Margutti:2010MNRAS.406.2149M}, suggesting that they may
share the same radiation mechanism as the prompt emission.

From a physical perspective, Salmonson et al.
\cite{Salmonson:2000ApJ...544L.115S} proposed that the $L_{\rm
p}-\tau_{\rm lag}$ correlation is purely a kinematic effect, where
the peak luminosity is proportional to the Lorentz factor $\Gamma$
of the jet shell, and the lag time is inversely proportional to
$\Gamma$. In this model, the luminosity in the comoving frame of
GRB emission region remains nearly constant, while the apparent
luminosity measured by the observer is determined entirely by the
Lorentz factor along the line of sight. Consequently, the Lorentz
factor and the observed luminosity must vary within the same order
of magnitude. This clearly does not align with the case of GRBs,
since the variation in peak luminosity far exceeds the range of
Lorentz factor variation derived from the afterglow data
\cite{Schaefer:2004ApJ...602..306S,Chen:2018MNRAS.478..749C,
Ghirlanda:2018A&A...609A.112G}.

Schaefer \cite{Schaefer:2004ApJ...602..306S} suggested that the
lag time partially reflects the pulse cooling process. In
high-luminosity GRBs, the cooling time is brief, resulting in a
shorter lag, whereas the cooling time of low-luminosity GRBs is
longer, leading to an extended lag. In the context of internal
shocks, Ioka et al. \cite{Ioka:2001ApJ...554L.163I} attributed the
variations in luminosity and spectral lag across different GRBs to
the observer's viewing angle. GRBs with high luminosity and short
lag are observed at smaller viewing angles, while low-luminosity
GRBs with longer lag are associated with larger viewing angles.
This interpretation also offers insight into the $L-V$
correlation. Additionally, the curvature effect may also
contribute to the observed spectral lag
\cite{Shen:2005MNRAS.362...59S,Lu:2006MNRAS.367..275L}.

\subsubsection{The $L_{\rm p}-V$ correlation}
Reichart et al. \cite{Reichart:2001ApJ...552...57R} analyzed a
sample of 20 GRBs with known redshifts. They found a positive
correlation between the peak luminosity ($L_{\rm p}$) and time
variability ($V$), i.e.,
\begin{eqnarray}
L_{\rm p} \propto V^{3.3^{+1.1}_{-0.9}}.
\end{eqnarray}
This correlation is consistent with the results reported by
Fenimore et al. \cite{Fenimore:2000astro.ph..4176F}, though
the definition of $V$ varies slightly
between the two studies. Guidorzi et al.
\cite{Guidorzi:2005MNRAS.363..315G} reanalyzed the $L_{\rm p}-V$
correlation using a larger sample of 32 GRBs. While the
correlation still holds, the power-law index differs significantly
from previous one. Reichart et al.
\cite{Reichart:2005astro.ph..8111R} argued that this discrepancy
stemmed from the use of improper statistical methods. They updated
the index as 3.4, which is consistent with earlier findings.

Different analysis method and different definition of variability
can result in various slopes in the $L_{\rm p}-V$ correlation
\cite{Guidorzi:2006MNRAS.371..843G,
Li:2006MNRAS.366..219L,Rizzuto:2007MNRAS.379..619R}. With a
substantial increase in sample size, Guidorzi et al.
\cite{Guidorzi:2024A&A...690A.261G} observed that the $L_{\rm
p}-V$ correlation became so scattered that it was challenging to
establish a clear relationship. Nevertheless, an inverse
correlation between the peak luminosity and the minimum variability
timescale was identified \cite{Camisasca:2023A&A...671A.112C},
despite the significant dispersion.

In the context of internal shocks, the observed $L-V$ relationship
may arise from the dynamics of relativistic jets in GRBs
\cite{Schaefer:2007ApJ...660...16S}. Both luminosity and
variability are governed by the Lorentz factor of the jet shell.
Owing to relativistic beaming effects, the size of the radiation
region observed by a distant observer is inversely proportional to
the Lorentz factor of the jet. Consequently, GRBs with higher
Lorentz factors exhibit greater luminosity and a smaller emission
region, resulting in increased variability.

This correlation may also be influenced by the opening angle of
the jet
\cite{Salmonson:2002ApJ...569..682S,Kobayashi:2002ApJ...577..302K}.
It is argued that jets with a smaller opening angle exhibit lower
baryon loading, potentially leading to a higher Lorentz factor.
Furthermore, the photospheric effect could play a role in the
$L-V$ correlation
\cite{Kobayashi:2002ApJ...577..302K,Meszaros:2002ApJ...578..812M,
Guidorzi:2024A&A...690A.261G}. In GRBs with a lower Lorentz
factor, the internal shock radius of the shell will be small and
may lie below the photospheric radius. This would smooth the pulse
profile of the prompt emission, thereby reducing the variability.
Salafia et al. \cite{Salafia:2016MNRAS.461.3607S} proposed that
the off-axis effect can significantly broaden the prompt pulse,
smearing out the light curve variability.

\subsubsection{The $E_{\rm p}-E_{\rm iso}$ correlation}

The $E_{\rm p}-E_{\rm iso}$ correlation, commonly known as the
Amati relation, is among the most discussed empirical correlations
in GRB studies. It was first introduced by Amati et al. in
analyzing 12 BeppoSAX GRBs with known redshifts
\cite{Amati:2002A&A...390...81A}, which is expressed as
\begin{eqnarray}
E_{\rm p} \propto E_{\rm iso}^{0.52 \pm 0.06}.
\end{eqnarray}
Here, the peak energy $E_{\rm p}$ is derived by fitting the
time-integrated spectrum using the Band function
\cite{Band:1993ApJ...413..281B}, while the isotropic equivalent
energy $E_{\rm iso}$ represents the total energy within the
$1-10^4$ keV range. It was further confirmed that X-ray rich GRBs
(XRRs) and X-ray flashes (XRFs) also follow the same Amati
relation as classical LGRBs
\cite{Lamb:2004NewAR..48..459L,Sakamoto:2004ApJ...602..875S}.

Later, Amati et al. \cite{Amati:2008MNRAS.391..577A} analyzed an
expanded sample of 70 GRBs and XRFs, and confirmed that the Amati
relation remains valid, with a slope consistent with previous
findings. With increasing sample sizes and improved data quality,
researchers have conducted more comprehensive studies of the Amati
relation
\cite{Amati:2012IJMPS..12...19A,Amati:2013IJMPD..2230028A,
Geng:2013ApJ...764...75G,Demianski:2017A&A...598A.112D}. It is
found that although the derived slopes vary slightly, they
generally cluster around 0.5.

Although SGRBs do not adhere to the Amati relation establised for
traditional LGRBs and XRFs, they occupy a parallel region in the
$E_{\rm p}-E_{\rm iso}$ plane \cite{Amati:2012IJMPS..12...19A}. This
indicates that, on average, SGRBs exhibit lower isotropic equivalent
energy at similar peak energies compared to LGRBs
\cite{Tsutsui:2008MNRAS.386L..33T}.

Nava et al. \cite{Nava:2012MNRAS.421.1256N} analyzed the Amati
relation based on \textit{Swift} GRBs across different redshift
bins. They found that the slope of the correlation does not evolve
with redshift. Demianski et al.
\cite{Demianski:2017A&A...598A.112D} parameterized the redshift
evolution of the observables in the Amati relation and confirmed
the absence of redshift evolution in this relation. Furthermore,
a similar relation has also been found to exist inside each GRB
based on time-resolved spectral analysis, with both the slope and
normalization being similar to those derived from time-integrated
spectrum
\cite{Basak:2012ApJ...749..132B,Basak:2013MNRAS.436.3082B}.

Despite extensive discussions regarding the Amati relation since
its discovery, the physical mechanism responsible for this
correlation remains poorly understood. This limited understanding
is largely due to our incomplete knowledge of the radiation
mechanisms driving prompt GRB emission, the processes involved in
electron acceleration by internal shocks, and the amplification of
magnetic fields in the radiation region
\cite{Parsotan:2022Univ....8..310P}. Photospheric emission is a
natural prediction of the standard fireball model for GRBs and may
influence the prompt emission spectrum
\cite{Thompson:2007ApJ...666.1012T,
Pe'er:2006ApJ...642..995P,Giannios:2007A&A...469....1G,Giannios:2012MNRAS.422.3092G}.
Thompson et al. \cite{Thompson:2007ApJ...666.1012T} interpreted
the thermally peaked emission, Doppler boosted by the GRB outflow,
as the peak energy, while the non-thermal high-energy emission
arises from inverse Compton scattering by the relativistic
electrons outside the photosphere. This model predicts a power-law
index of 0.5 for the Amati relation.

Eichler $\&$ Levinson \cite{Eichler:2004ApJ...614L..13E} proposed
that a ring-shaped fireball with viewing angle effects could
reproduce the Amati relation, with a power-law index of 0.5.
Without specifying radiation processes, Dado et al.
\cite{Dado:2012ApJ...749..100D} suggested that the Amati relation
may have a dynamical origin and they derived a slope of 0.5 based
on the so-called cannonball model. Additionally, several authors
have attempted to reproduce the Amati relation through numerical
simulations
\cite{Yamazaki:2004ApJ...606L..33Y,Kocevski:2012ApJ...747..146K,
Mochkovitch:2015A&A...577A..31M}. Mochkovitch et al.
\cite{Mochkovitch:2015A&A...577A..31M} demonstrated that the
internal shock model can explain the observations. They further
provided strict constraints on the dynamics and the microphysical
parameters governing shock energy dissipation.

Some GRBs deviate significantly from the Amati relation,
indicating that they could be off-axis events
\cite{Heussaff:2013A&A...557A.100H,
Stanway:2015MNRAS.446.3911S,D'Elia:2018A&A...619A..66D,Farinelli:2021MNRAS.501.5723F}.
Xu et al. \cite{Xu:2023A&A...673A..20X} analytically derived the
on-axis and off-axis Amati and Yonetoku relations (also see
Section \ref{sec2.1.4}) within the framework of the standard
fireball model, based on the optically thin synchrotron radiation
mechanism under the fast cooling scenario. In the on-axis cases,
the Amati relation follows $E_{\rm p} \propto E_{\rm iso}^{0.5}$,
whereas in the off-axis cases, this correlation is modified to
$E_{\rm p} \propto E_{\rm iso}^{1/4 \sim 4/13}$, where the range
reflects variations in the viewing angle beyond the jet core.

\subsubsection{The $E_{\rm p}-L_{\rm p}$ correlation}
\label{sec2.1.4}

Yonetoku et al. \cite{Yonetoku:2004ApJ...609..935Y} found a
positive correlation between $E_{\rm p}$ and $L_{\rm p}$, known as
the Yonetoku relation, which is expressed as
\begin{eqnarray}
\frac{L_{\rm p}}{10^{52}\, \mathrm{erg \, s^{-1}}} = \left( 2.34^{+2.29}_{-1.76}
\right) \times 10^{-5} \left[ \frac{E_{\rm p} (1+z)}{1 \, \mathrm{keV}} \right]
^{2.0 \pm 0.2},
\end{eqnarray}
where the peak energy $E_{\rm p}$ is derived from the Band
function \cite{Band:1993ApJ...413..281B}, and the peak luminosity
$L_{\rm p}$ is the 1-second peak luminosity. Using this relation,
Yonetoku et al. \cite{Yonetoku:2004ApJ...609..935Y} calculated the
pseudo-redshifts of 689 BATSE GRBs and argued that there is redshift
evolution in the relation.

Further research by Ghirlanda et al.
\cite{Ghirlanda:2004A&A...422L..55G} revealed that BATSE SGRBs
without measured redshifts adhere to the Yonetoku relation but
deviate from the Amati relation. If SGRBs follow the Yonetoku
relation of LGRBs, their pseudo-redshift distribution resembles
that of LGRBs, albeit with a lower average redshift. The validity
of the Yonetoku relation in SGRBs with confirmed redshifts has
been verified by a number of studies
\cite{Ghirlanda:2009A&A...496..585G,Zhang:2009ApJ...703.1696Z,
Zhang:2012ApJ...755...55Z,Shahmoradi:2015MNRAS.451..126S,D'Avanzo:2014MNRAS.442.2342D,
Tsutsui:2013MNRAS.431.1398T}. In the $E_{\rm p}-L_{\rm p}$ plane,
SGRBs are dimmer by a factor of approximately 5 compared to LGRBs
with same $E_{\rm p}$ \cite{Tsutsui:2013MNRAS.431.1398T}. Some
low-luminosity GRBs clearly deviate from the Yonetoku relation,
suggesting an origin possibly different from that of classical
LGRBs \cite{Yonetoku:2010PASJ...62.1495Y}.

Yonetoku et al. \cite{Yonetoku:2010PASJ...62.1495Y} utilized an expanded sample consisting
of 101 GRBs with well-defined spectral information to update the Yonetoku relation as
\begin{eqnarray}
L_{\rm p} = 10^{52.43 \pm 0.037} \times \left[ \frac{E_{\rm p} (1+z)}{355 \, \mathrm{keV}}
\right]^{1.60 \pm 0.082},
\end{eqnarray}
where the slope of the updated relation is slightly smaller than the original one but
remains consistent at the $2\sigma$ confidence level.

Some authors have argued that the Yonetoku and Amati relations
could be significantly impacted by detector threshold truncation
effects
\cite{Nakar:2005MNRAS.360L..73N,Band:2005ApJ...627..319B,Butler:2007ApJ...671..656B},
casting doubts on their authenticity. However, subsequent detailed
analyses indicated that both relations are intrinsic
\cite{Yonetoku:2010PASJ...62.1495Y}. The Yonetoku relation shows
slight redshift evolution, whereas the Amati relation is modestly
affected by detector threshold effects, which may contribute to
the observed dispersion in the Amati relation. Yonetoku et al.
\cite{Yonetoku:2004ApJ...609..935Y} proposed that the redshift
dependence observed in the peak luminosity could arise from the
evolution of GRB progenitors or jet opening angles with redshift.

According to the standard fireball model with synchrotron
radiation \cite{Xu:2023A&A...673A..20X}, the on-axis Yonetoku
relation can be derived analytically as $E_{\rm p} \propto L_{\rm
p}^{0.5}$, in which the power-law index is largely consistent with
the observations. In the off-axis cases, the Yonetoku relation is
modified to $E_{\rm p} \propto L_{\rm p}^{1/6 \sim 4/17}$, where
the power-law index is flatter than that of the Amati relation.
The outliers of the Amati and Yonetoku relations could stem from
viewing angle effect. Interestingly, they are more easily
identifiable in the Yonetoku relation than in the Amati relation.

While the optically thin synchrotron radiation model can
effectively account for the spectral-luminosity (or energy)
correlations, its simplified parameterization and significant
uncertainties in various physical parameters limit the predictive
power over emission properties. Consequently, the photosphere
model, based on thermal radiation, has gained favor among some
researchers and has been applied to explain the Yonetoku relation
\cite{Giannios:2012MNRAS.422.3092G,
Ito:2014ApJ...789..159I,Ito:2021ApJ...918...59I,Ito:2024ApJ...961..243I,
Fan:2012ApJ...755L...6F}. Ito et al.
\cite{Ito:2019NatCo..10.1504I} conducted three-dimensional
hydrodynamic simulations, demonstrating that the Yonetoku relation
naturally arises from viewing angle effects. The dispersion in the
Yonetoku relation may stem from the intrinsic properties of GRB
jets, such as the jet power, Lorentz factor, and duration.

\subsection{prompt and afterglow correlations}

\subsubsection{The $L_{\rm X}-T_{\rm a}$ correlation}

Dainotti et al. \cite{Dainotti:2008MNRAS.391L..79D} analyzed 33
LGRBs and discovered an anti-correlation involving the X-ray
afterglow plateau of GRBs, known as the Dainotti relation, which
has the following form,
\begin{eqnarray}
\log L_{\rm X} = a + b\log T_{\rm a},
\end{eqnarray}
where $T_{\rm a}$ represents the end time of the plateau phase in
the GRB rest frame (in seconds), and $L_{\rm X}$ is the
corresponding X-ray luminosity at $T_{\rm a}$ (in $\rm erg~ \rm
s^{-1}$). The $K$-correction accounting for the cosmic expansion
is applied in calculating the X-ray luminosity
\cite{Bloom:2001AJ....121.2879B}. The afterglow light curve is
described using a phenomenological two-component model
\cite{Willingale:2007ApJ...662.1093W}. The normalization constant
$a$ and slope $b$ in the original version are 48.54 and
$-0.74^{+0.20}_{-0.19}$, respectively. Subsequently, Dainotti et
al. \cite{Dainotti:2010ApJ...722L.215D} employed an expanded
sample of 62 LGRBs to update the Dainotti relation as $L_{\rm
X}\propto T_{\rm a}^{-1.06^{+0.27}_{-0.28}}$. This relation also
holds for intermediate-duration GRBs, albeit with a steeper slope.
The Dainotti relation remains valid for larger samples, and
although the slopes reported in various studies may differ
slightly, they are largely consistent.
\cite{Cardone:2010MNRAS.408.1181C,Dainotti:2010ApJ...722L.215D,
Sultana:2012ApJ...758...32S,Bernardini:2012A&A...539A...3B,Mangano:2012MSAIS..21..143M,
Dainotti:2013ApJ...774..157D,Postnikov:2014ApJ...783..126P,Dainotti:2017A&A...600A..98D}.

The varying slopes derived from different samples may be
influenced by redshift evolution and selection bias, which could
undermine the robustness of the correlation as a cosmological
probe. Dainotti et al. \cite{Dainotti:2013ApJ...774..157D} applied
the Efron $\&$ Petrosian (EP) method
\cite{Efron:1992ApJ...399..345E} to correct for the redshift
evolution of $L_{\rm X}$ and $T_{\rm a}$. Using a sample of 101
GRBs with well-defined light curves, they demonstrated that the
$L_{\rm X}-T_{\rm a}$ correlation is intrinsic for a significant
portion of GRBs, with a slope of $-1.07$. Moreover, similar 2D
Dainotti relation has been investigated in the optical and radio
bands
\cite{Dainotti:2020ApJ...905L..26D,Levine:2022ApJ...925...15L}.

Dainotti et al. \cite{Dainotti:2017A&A...600A..98D} identified a
significant difference in the Dainotti relation among subsamples
of LGRBs with and without supernovae. This difference persisted
even after applying the EP method to correct for the redshift
evolution in LGRBs without supernovae.

Additionally, Dainotti et al. \cite{Dainotti:2016ApJ...825L..20D}
combined the $L_{\rm X}-T_{\rm a}$ correlation with the $L_{\rm p}$
of the GRB prompt phase to derive a new three-parameter relation,
denoted as
\begin{eqnarray}
\log L_{\rm X} = (15.75 \pm 5.3) - (0.77 \pm 0.1) \log T_{\rm a} +
(0.67 \pm 0.1) \log L_{\rm p}.
\end{eqnarray}
Following the convention of Dainotti et al.
\cite{Dainotti:2016ApJ...825L..20D,Dainotti:2017ApJ...848...88D},
we refer to the $L_{\rm X}-T_{\rm a}-L_{\rm p}$ correlation as the
3D Dainotti relation. Dainotti et al.
\cite{Dainotti:2016ApJ...825L..20D} focused on GRBs with
sufficient data points in the plateau phase to accurately describe
their light curves, designating them as the ``gold'' sample. The
intrinsic dispersion of the 3D Dainotti relation derived based on
the gold sample is $\sigma_{\rm int} = 0.27 \pm 0.04$,
significantly smaller than that of the original 2D Dainotti
relation. Further study has demonstrated that the 3D Dainotti
relation applies to all classes of GRBs, with the intrinsic
dispersion reduced to $\sigma_{\rm int} = 0.22 \pm 0.1$ when using
platinum sample selected under stringent criteria
\cite{Dainotti:2020ApJ...904...97D}.

The X-ray plateau phase is not a phenomenon predicted by the traditional
forward shock model \cite{Zhang:2006ApJ...642..354Z}. It may be linked
to the reactivation of the central engine or reflect the angular structure
of the GRB jet
\cite{Yi:2022ApJ...924...69Y,Salafia:2016MNRAS.461.3607S,Cannizzo:2009ApJ...700.1047C},
both of which are crucial for understanding the nature of GRBs.

In a framework involving a black hole with fallback accretion,
Cannizzo et al. \cite{Cannizzo:2011ApJ...734...35C} proposed that
the Dainotti relation can be reproduced if the accreted matter has
a unified mass. The power-law index in the Dainotti relation is
approximately $-1$, indicating that a constant energy reservoir
may be involved in the plateau phase. A plausible explanation is
that the central engine of the GRB is not a black hole, but a
millisecond magnetar
\cite{Dai:1998A&A...333L..87D,Zhang:2001ApJ...552L..35Z,Dall'Osso:2011A&A...526A.121D,
Geng:2016ApJ...825..107G}, whose spin-down energy is injected into
the external forward shock, producing the observed X-ray plateau.
Rowlinson et al. \cite{Rowlinson:2014MNRAS.443.1779R} derived a
scaling relationship between the plateau luminosity and its
duration under the magnetar energy injection scenario, i.e.,
\begin{eqnarray}
\log L_{\rm X} \sim \log(10^{52} I_{45}^{-1} P_{0,-3}^{-2}) - \log T_{\rm a},
\end{eqnarray}
where $I_{45}$ is the moment of inertia of the magnetar in units
of $10^{45}~\rm{g~\rm {cm}^2}$ and $P_{0,-3}$ represents the
initial spin period of the magnetar in units of millisecond. This
is consistent with the anti-correlation identified by Dainotti et
al. \cite{Dainotti:2013ApJ...774..157D}, i.e.,
$L_{\rm X}\propto T_{\rm a}^{-1}$. Variations in the
initial spin period of the magnetar may contribute to the
dispersion observed in the Dainotti relation. However, while the
energy injection scenario often predicts achromatic brightening
across multiple bands \cite{Geng:2013ApJ...779...28G}, the
expected brightening is not observed in the afterglow light curves
in other bands.

An alternative explanation is that the GRB jets may possess
angular structures rather than the simplified top-hat shape with
sharp edges \cite{Salafia:2016MNRAS.461.3607S,
Beniamini:2020MNRAS.492.2847B,Beniamini:2022MNRAS.515..555B,O'Connor:2023SciA....9I1405O,
Oganesyan:2020ApJ...893...88O}. The shallow decay phase can
naturally result from off-axis observations of forward shock
radiation emitted by structured jets propagating through the
circumburst medium \cite{Beniamini:2020MNRAS.492.2847B}. This
model has the advantage of not requiring long-term central engine
activities and can produce plateaus lasting for $10^2$ to $10^5$
seconds within the appropriate parameter space, in both
interstellar medium and wind environments.

\subsubsection{The $L_{\rm X}-T_{\rm a}-E_{\rm iso}$ correlation}

Xu $\&$ Huang \cite{Xu:2012A&A...538A.134X} collected 77 GRBs that
exhibited plateaus in their X-ray afterglows and introduced the
$E_{\rm iso}$ of the prompt phase as the third parameter. They derived
a tight three-parameter correlation, i.e., the
$L_{\rm X}-T_{\rm a}-E_{\rm iso}$ correlation, as,
\begin{eqnarray}
\log \left( \frac{L_{\rm X}}{10^{47}~\text{erg}~\text{s}^{-1}} \right) = a + b \times
\log \left( \frac{T_{\rm a}}{10^3 ~ \text{s}} \right) + c\times \log
\left( \frac{E_{\text{iso}}}{10^{53}~ \text{erg}} \right),
\end{eqnarray}
where a, b, and c are coefficient obtained by fitting the observational data.
The $K$-correlation is also considered in the calculation of $E_{\rm iso}$. The best
fitting results derived from the so-called Markov-chain Monte Carlo (MCMC) techniques
are $a = 0.81 \pm 0.07$, $b = -0.91 \pm 0.09$, $c = 0.59 \pm 0.05$, and
$\sigma_{\rm int} = 1.15 \pm 0.12$, where $\sigma_{\rm int}$ represents the intrinsic
dispersion of the correlation. A smaller $\sigma_{\rm int}$ indicates a tighter
correlation. After applying additional selection criteria, Xu $\&$ Huang
\cite{Xu:2012A&A...538A.134X} defined a golden sample consisting of 47 LGRBs and 8
intermediate-duration GRBs. The best fitting results for this refined sample
are $a = 1.17 \pm 0.09$, $b = -0.87 \pm 0.09$, $c = 0.88 \pm 0.08$, and
$\sigma_{\rm int} = 0.43 \pm 0.05$. Evidently, after rigorous sample selection,
the intrinsic dispersion is significantly reduced, leading to a tighter
$L_{\rm X}-T_{\rm a}-E_{\rm iso}$ correlation.

\begin{figure}[H]
\includegraphics[width=10.5 cm]{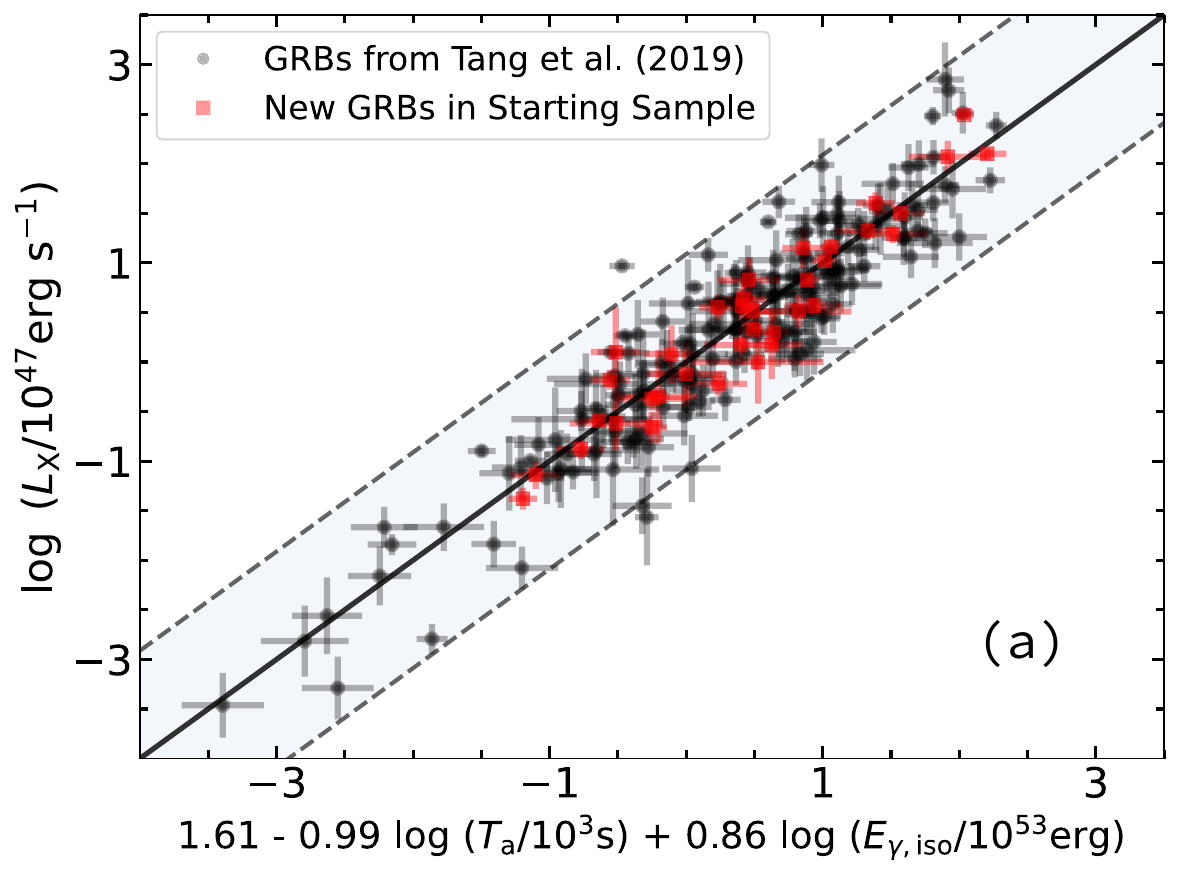}
\includegraphics[width=10.5 cm]{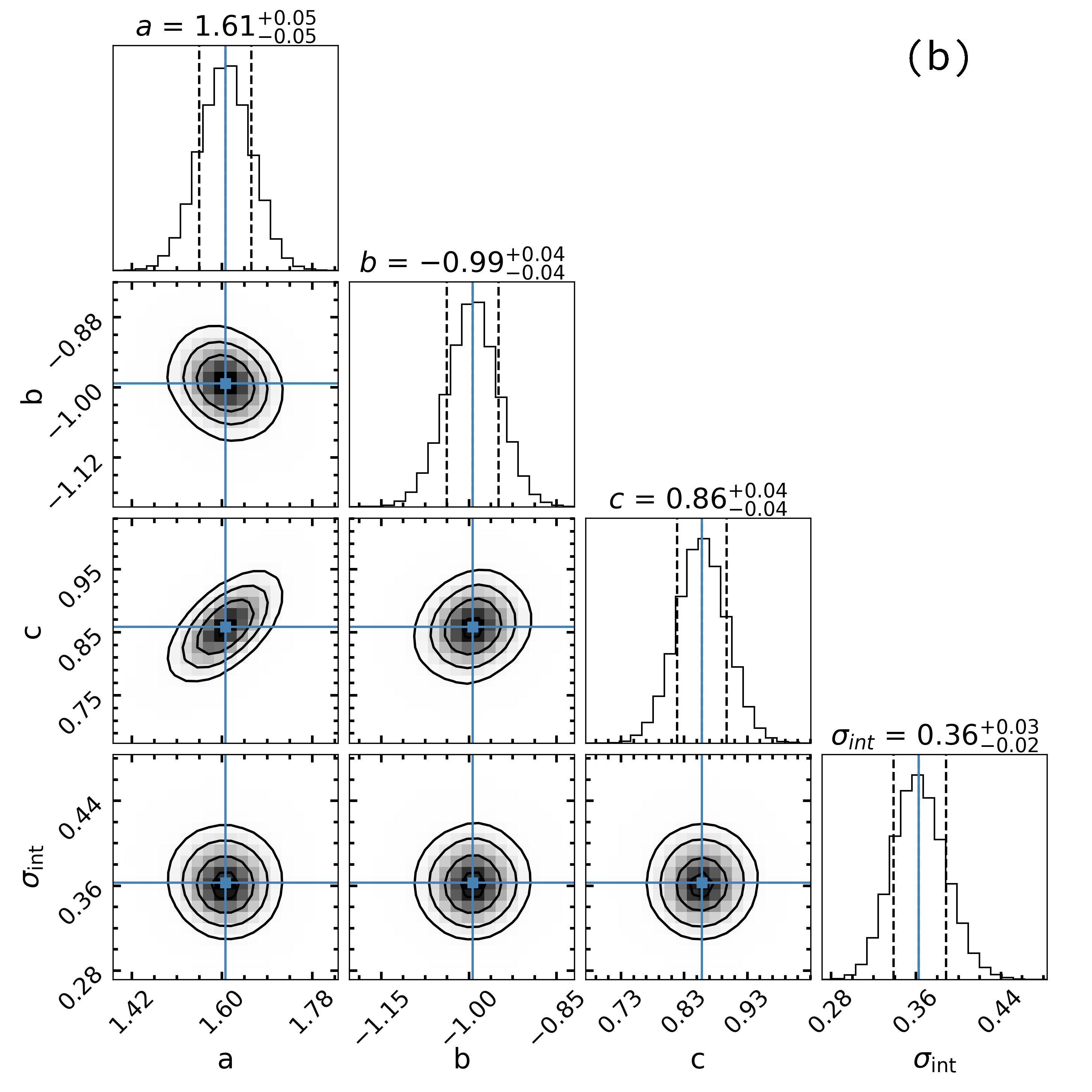}
\caption{(a) The $L_{\rm X}-T_{\rm a}-E_{\rm iso}$ correlation
updated with 210 GRBs by Deng et al. \cite{Deng:2023ApJ...943..126D}.
The solid line and dashed lines denote the best-fit result and
3$\sigma$ confidence level, respectively. The 174 GRBs from
Tang et al. \cite{Tang:2019ApJS..245....1T} are represented by
black dots, and the red squares indicate the 36 new GRBs included
in the analysis of Deng et al. \cite{Deng:2023ApJ...943..126D}.
(b) The posterior probability distributions of the fitting parameters
in the $L_{\rm X}-T_{\rm a}-E_{\rm iso}$ correlation from Deng et al.
\cite{Deng:2023ApJ...943..126D}. The vertical solid
lines and dashed lines represent the mean values and 1$\sigma$ confidence
level, respectively.
\label{fig1}}
\end{figure}

Tang et al. \cite{Tang:2019ApJS..245....1T} analyzed a larger
sample of 174 plateau GRBs with well-defined light curves and
known redshifts. They confirmed the existence of the $L_{\rm
X}-T_{\rm a}-E_{\rm iso}$ correlation and updated it as $L_{\rm X}
\propto T_{\rm a}^{-1.01 \pm 0.05} E_{\rm iso}^{0.84 \pm 0.04}$.
It should be noted that Tang et al.
\cite{Tang:2019ApJS..245....1T} employed a smooth broken power-law
function to fit the afterglow light curve, whereas Dainotti et al.
\cite{Dainotti:2008MNRAS.391L..79D} used a two-component
phenomenological model proposed by Willingale et
al. \cite{Willingale:2007ApJ...662.1093W}. The difference in the
method may lead to slight discrepancies in the fitted plateau
duration and the corresponding X-ray luminosity. Deng et al.
\cite{Deng:2023ApJ...943..126D} expanded the sample of Tang et al.
\cite{Tang:2019ApJS..245....1T} with additional 36 plateau GRBs,
updating the $L_{\rm X}-T_{\rm a}-E_{\rm iso}$ correlation as
$L_{\rm X} \propto T_{\rm a}^{-0.99 \pm 0.04} E_{\rm iso}^{0.86
\pm 0.04}$, with a slight reduction in $\sigma_{\rm int}$ from
$0.39 \pm 0.03$ to $0.36 \pm 0.03$.
The best-fit $L_{\rm X}-T_{\rm a}-E_{\rm iso}$ correlation and the
constrains on the coefficients are displayed in Figure \ref{fig1}.

From a physical perspective, the $L_{\rm X}-T_{\rm a}-E_{\rm iso}$
correlation may indicate that the X-ray plateaus originate from
central engine energy injection, akin to discussions surrounding
the $L_{\rm X}-T_{\rm a}$ correlation. Xu et al.
\cite{Xu:2012A&A...538A.134X} proposed that the injected energy
from the magnetar may manifest as relativistic positron-electron
wind
\cite{Dai:2004ApJ...606.1000D,Yu:2007A&A...470..119Y,Yu:2007ApJ...671..637Y},
a scenario similar to that employed by Geng et al.
\cite{Geng:2016ApJ...825..107G,Geng:2018ApJ...856L..33G} to explain the late-time optical
brightening observed in some GRBs. As the positron-electron wind
interacts with the forward shock, it induces a long-duration
reverse shock that propagates through the wind itself,
contributing to the shallow decay phase observed in the X-ray
afterglow. A key advantage of this model is its explanation of the
chromatic break observed in the afterglow phase, achieved by the
combined contributions of multiple emission regions to the total
flux. The X-ray plateau luminosity reflects the rate of energy
injection from the central engine. Differences in the magnetic
field strength and spin period of the nascent magnetar may account
for the observed dispersion within this correlation. Dai et al.
\cite{Dai:2004ApJ...606.1000D} demonstrated that the energy
injected by the magnetar into the forward shock, which produces
the observable plateau, should be comparable to the energy of the
initial fireball that drives the prompt radiation. This is largely
consistent with the $L_{\rm X}-T_{\rm a}-E_{\rm iso}$ correlation,
which essentially gives $L_{\rm X}T_{\rm a} \propto E_{\rm
iso}^{0.86}$.

\subsubsection{The $E_{\rm X,iso}-E_{\rm iso}-E_{\rm p}$ correlation}

Bernardini et al. \cite{Bernardini:2012MNRAS.425.1199B} discovered a
tight three-parameter correlation in \textit{Swift} GRBs involving the
$E_{\rm iso}$ in the 1-10$^4$ keV range, the peak energy $E_{\rm p}$,
and the isotropic equivalent energy $E_{\rm X,iso}$ of X-ray afterglow
in the 0.3-10 keV band. i.e.,
\begin{eqnarray}
\log \left( \frac{E_{\rm X,\text{iso}}}{\text{erg}} \right) = (1.06 \pm 0.06) \log
\left( \frac{E_{\text{iso}}}{\text{erg}} \right)
- (0.74 \pm 0.10) \log \left( \frac{E_{\text{p}}}{\text{keV}} \right) - (2.36 \pm 0.25).
\end{eqnarray}
The $E_{\rm X,iso}-E_{\rm iso}-E_{\rm p}$ correlation was further
examined in depth by Margutti et al.
\cite{Margutti:2013MNRAS.428..729M}, wherein various observational
correlations involving both prompt and afterglow parameters were
also investigated. This correlation can be interpreted as arising
from the inclusion of $E_{\rm X, iso}$ as a third variable into
the Amati relation, which reduces the intrinsic dispersion of the
original relation. Notably, while SGRBs do not satisfy the Amati
relation derived from LGRBs, both types of GRBs conform to the
same $E_{\rm X,iso}-E_{\rm iso}-E_{\rm p}$ correlation. This
indicates long and short GRBs may share similar characteristics,
offering insights into the underlying physics of both groups in a
unified picture \cite{Margutti:2013MNRAS.428..729M}. Subsequently,
Zaninoni et al. \cite{Zaninoni:2016MNRAS.455.1375Z} validated the
$E_{\rm X,iso}-E_{\rm iso}-E_{\rm p}$ correlation with an expanded
sample. Although long and short GRBs adhere to the same $E_{\rm
X,iso}-E_{\rm iso}-E_{\rm p}$ correlation, they occupy separate
regions in the 2D correlation plane. Furthermore, Zaninoni et al.
\cite{Zaninoni:2016MNRAS.455.1375Z} found that an intermediate
group of GRBs exists between long and short GRBs in this 2D plane,
which encompasses ultral-long GRBs and low-energy GRBs.

A two-parameter correlation involving $E_{\rm p}$ and $\epsilon$
serves as a convenient proxy for the $E_{\rm X,iso}-E_{\rm
iso}-E_{\rm p}$ correlation, where $\epsilon$, defined as
$\epsilon=E_{\rm X,iso}/E_{\rm iso}$, inversely represents the
emission efficiency of GRBs. Bernardini et al.
\cite{Bernardini:2012MNRAS.425.1199B} found that $\epsilon \propto
E_{\rm p}^{-0.58}$. This finding suggests that the $E_{\rm
X,iso}-E_{\rm iso}-E_{\rm p}$ correlation may be connected with
the Lorentz factor of the GRB jet, as a faster outflow converts
more energy into the prompt phase. Additionally, the photospheric
model provides a natural explanation for the $E_{\rm X,iso}-E_{\rm
iso}-E_{\rm p}$ correlation, since brighter GRBs exhibit a higher
radiation efficiency \cite{Lazzati:2013ApJ...765..103L}. The
cannonball model has also been proposed as an alternative
explanation for this correlation \cite{Dado:2013ApJ...775...16D}.

\subsubsection{The Combo relation}

Izzo et al. \cite{Izzo:2015A&A...582A.115I} analyzed a sample of
60 LGRBs. They combined the Amati relation with the $E_{\rm
X,iso}-E_{\rm iso}-E_{\rm p}$ correlation to derive a tight
four-parameter correlation that involves both the prompt emission
parameters and the afterglow parameters. It is called the Combo
relation, which can be expressed as
\begin{eqnarray}
\log \left(\frac{L_0}{\text{erg/s}} \right) = \log \left( \frac{A}{\text{erg/s}} \right) +
\gamma \left( \log \left( \frac{E_{\rm p}}{\text{keV}} \right) - \frac{1}{\gamma} \log
\left( \frac{\tau/\text{s}}{|1 + \alpha_X|} \right) \right),
\end{eqnarray}
where $L_0$, $\tau$ and $\alpha_X$ are the intial X-ray luminosity
of the plateau phase, the end time of the plateau phase, and
power-law decay index post the shallow phase, respectively. A
phenomenological light curve expression, $L(t) =
L_0(1+t/{\tau})^{{\alpha}_X}$ as proposed by Ruffini et al.
\cite{Ruffini:2014A&A...565L..10R}, is adopted to fit the X-ray
data and further derive the constant $L_0$, which is crucial for
determining the Combo relation. The best fit parameters provided
by Izzo et al. \cite{Izzo:2015A&A...582A.115I} are
$\log[A/(\text{erg/s})]=49.94 \pm 0.27$, $\gamma = 0.74 \pm 0.1$,
and $\sigma_{\rm int} = 0.33 \pm 0.04$. Muccino et al.
\cite{Muccino:2021ApJ...908..181M} later confirmed the validity of
this correlation with an expanded sample of 174 GRBs, updating the
coefficients as $\log[A/(\text{erg/s})] = 49.66 \pm 0.2$, $\gamma
= 0.84 \pm 0.08$, and $\sigma_{\rm int} = 0.37 \pm 0.02$. The
slope parameter $\gamma$ remains largely constant across different
redshift bins, indicating that the Combo relation does not evolve
at different redshift.

Theoretically, the Combo relation reflects the connection between
the prompt radiation and the X-ray afterglow, as well as the
characteristics of GRB outflow and the surrounding environment
\cite{Muccino:2017MNRAS.468..570M}. Muccino et al.
\cite{Muccino:2017MNRAS.468..570M} used an external forward shock
model, which accelerates electrons into a power-law distribution,
to explain the physical origin of the Combo relation. In this
framework, the parameters such as $E_{\rm p}$, $L_0$, and $\tau$
depend on the initial Lorentz factor of the GRB jet, and the
different spectral energy distribution of the emitting source lead
to the dispersion of the Combo relation.

\section{Application of the Observed Correlations}

\label{sec3}

\subsection{Usage of the observed correlation in cosmology}

An essential preliminary step in using GRBs to constrain
cosmological model parameters is deriving the Hubble diagram of
GRBs based on known empirical correlations. The Hubble diagram
represents the relationship between redshift $z$ and the distance
modulus $\mu$, defined as
\begin{eqnarray}
\mu = 5 \log \frac{d_L}{\text{Mpc}} + 25 = 5 \log
\frac{d_L}{\text{cm}} - 97.45.
\end{eqnarray}
Here $d_{L}$ denotes the luminosity distance, which is expressed
as
\begin{eqnarray}
d_L = (1 + z) \frac{C}{H_0} \begin{cases}
\frac{1}{\sqrt{\Omega_k}} \sinh \left[ \sqrt{\Omega_k} \int_0^z \frac{dz'}
{\sqrt{E(z', \Omega_m, \Omega_\Lambda, w(z'))}} \right], & \Omega_k > 0, \\[10pt]
\int_0^z \frac{dz'}{\sqrt{E(z', \Omega_m, \Omega_\Lambda, w(z'))}}, & \Omega_k = 0, \\[10pt]
\frac{1}{\sqrt{|\Omega_k|}} \sin \left[ \sqrt{|\Omega_k|} \int_0^z
\frac{dz'} {\sqrt{E(z', \Omega_m, \Omega_\Lambda, w(z'))}}
\right], & \Omega_k < 0,
\end{cases}
\end{eqnarray}
where $\Omega_{\rm m}$, $\Omega_{\rm k}$, and $\Omega_{\Lambda}$
are the matter density parameter, space curvature, and dark energy
density parameter, respectively. $C$ and $H_0$ represent the speed
of light and the Hubble constant, respectively. The specific form
of $E(z, \Omega_m, \Omega_\Lambda, w(z))$ depends on the
cosmological model and is generally defined as
\begin{eqnarray}
\begin{aligned}
E(z, \Omega_m, \Omega_\Lambda, w(z)) = \Omega_{\rm m}(1+z)^3 + \Omega_k(1+z)^2 \\
    + \Omega_{\Lambda}e^{3\int_{0}^{\ln(1+z)}1+\omega(z')d\ln(1+z')},
\end{aligned}
\end{eqnarray}
where $\omega (z)$ denotes the equation of state (EOS) parameter
of dark energy. In a parameterized CPL model \cite{Chevallier:2001IJMPD..10..213C,Linder:2003PhRvL..90i1301L}, $\omega (z) =
\omega_0 + \omega_a\frac{z}{1+z}$, representing the evolution of
the dark energy EOS with redshift. When $\omega_a$ is set to 0 and
$\omega_0$ varies freely, we call it the $\omega$CDM model. If
$\omega_0$ is further fixed at $-1$, the model reduces to the
standard $\Lambda$CDM model. Theoretically, one can identify the
cosmological model and determine its parameters by fitting the
observed GRB data.

\subsubsection{Cosmology with the Amati relation}

Amati et al. \cite{Amati:2008MNRAS.391..577A} utilized the Amati
relation derived from a sample of 70 LGRBs and XRFs to constrain
the $\Lambda$CDM cosmological parameters. Their results indicate
that the matter density parameter $\Omega_{\rm m}$ lies in a range
of 0.05 -- 0.40 for the flat assumption, and in 0.04 -- 0.50 for
the non-flat case, where the ranges correspond to the 1$\sigma$
confidence level. Numerical simulations indicate that the
uncertainty of cosmological parameters can be further reduced as
the GRB sample size increases. Wang et al.
\cite{Wang:2016A&A...585A..68W} compiled a sample of 151 LGRBs and
calibrated the Amati relation using SNe Ia. The cosmology
parameters derived by them are consistent with those derived based
on SNe Ia observations, yielding $\Omega_{\rm m}=0.271 \pm 0.019$
for the flat $\Lambda$CDM universe, and $\Omega_{\rm m}=0.225 \pm
0.044$ and $\Omega_{\Lambda}=0.640 \pm 0.082$ for the non-flat
$\Lambda$CDM model. The precision of $\Omega_{\rm m}$ was further
enhanced by combining 221 LGRBs with the Pantheon SNe Ia sample
\cite{Scolnic:2018ApJ...859..101S}, providing the best-fit
parameters of $\Omega_{\rm m}=0.29 \pm 0.01$ for the flat
$\Lambda$CDM model, and $\Omega_{\rm m}=0.34 \pm 0.04$ and
$\Omega_{\Lambda}=0.79 \pm 0.06$ for the non-flat $\Lambda$CDM
scenario \cite{Jia:2022MNRAS.516.2575J}.

Due to the limited observational data for low-redshift
GRBs (e.g., $z<0.1$), establishing the empirical correlations
depends on a pre-assumed cosmological model to derive isotropic
equivalent energy or luminosity. When such a relation is directly
applied for cosmology studies, the derived cosmological parameters
become dependent on the prior cosmological model, known as the
circularity problem \cite{Kodama:2008MNRAS.391L...1K}. A commonly
employed method to address this circularity problem involves
standardizing GRBs using SNe Ia data, as previously mentioned.
However, the results derived from this approach should be
interpreted with caution. This is because the systematics of SNe
Ia could bias the GRB Hubbble diagram, thereby amplifying the
uncertainties associated with using GRBs as a cosmological probe.
It is worth noting that Amati et al.
\cite{Amati:2013IJMPD..2230028A} constrained $\Omega_m$ as $\sim
0.3$ by using GRB data alone, without requiring calibration with
SNe Ia.

Moreover, the use of observational Hubble data (OHD)
enables the calibration of the GRB luminosity correlation in a
model-independent way
\cite{Ryan:2018MNRAS.480..759R,Capozziello:2018MNRAS.476.3924C,Yu:2018ApJ...856....3Y}.
The application of B\'ezier parametric curves
\cite{Amati:2019MNRAS.486L..46A,Luongo:2021Galax...9...77L} and
Gaussian process methods
\cite{Seikel:2012JCAP...06..036S,Wang:2022ApJ...926..178W,Hu:2021MNRAS.507..730H}
to fit the OHD is well-established in the literature. However,
calibrating GRBs using external calibrators such as SNe Ia and OHD
compromises their status as independent cosmological probes. In
addition, the calibration process is typically performed at low
redshift and subsequently extended to higher redshift regions. The
reliability of such calibration methods depends on the assumption
that the GRB luminosity correlations do not evolve with redshift.
Redshift evolution may arise from selection biases and
instrumental detection thresholds
\cite{Amati:2013IJMPD..2230028A,Dainotti:2018PASP..130e1001D,Dainotti:2017NewAR..77...23D},
making it crucial to assess its impact on the application of the
luminosity correlation in cosmology. Taking the Amati relation as
an example, although there is ongoing debate about the existence
of redshift evolution, its impact on the robustness of the Amati
relation across different redshifts is trivial
\cite{Nava:2012MNRAS.421.1256N,Amati:2013IJMPD..2230028A,Amati:2019MNRAS.486L..46A,Demianski:2017A&A...598A.112D}.
Ghirlanda et al. \cite{Ghirlanda:2010A&A...511A..43G} demonstrated
that both the normalization factor and slope of the Amati relation
are independent of redshift. Demianski et al.
\cite{Demianski:2017A&A...598A.112D} divided a sample of 162 LGRBs
into low- and high-redshift groups, with a redshift of 2 as the
boundary. They showed that the constrained results of the
$\Lambda$CDM and CPL models for the two groups were consistent
within 1 $\sigma$ error.

Other promising approaches to address the circularity
problem involve performing self-calibration within a narrow
redshift bin, or simultaneously fitting the parameters for the
luminosity correlation and cosmological models, without relying on
a background cosmological model. The quality of GRB samples is
critical for establishing GRBs as reliable independent
cosmological probes. Khadka et al.
\cite{Khadka:2020MNRAS.499..391K} performed a comprehensive
analysis of six cosmological models using the Amati relation,
applied to a meticulously compiled sample of 119 GRBs through the
simultaneous fitting method. Their findings confirmed that the
parameters of the Amati relation are model-independent,
underscoring its effectiveness in standardizing GRBs as reliable
cosmological probes. This represented the first demonstration of
such a result for GRBs. After excluding one GRB with insecure
redshift measurement, the remaining 118 GRBs, i.e., A118 dataset
\cite{Khadka:2020MNRAS.499..391K,Khadka:2021JCAP...09..042K,Liu:2022ApJ...931...50L,
FanaDirirsa:2019ApJ...887...13F,Cao:2024JCAP...10..093C,Cao:2023PhRvD.107j3521C},
exhibited low intrinsic dispersion and formed a consistent Amati
relation GRB sample \cite{Khadka:2021JCAP...09..042K}.
Additionally, Cao et al.
\cite{Cao:2021MNRAS.501.1520C,Cao:2022MNRAS.513.5686C,Cao:2023PhRvD.107j3521C}
conducted robust joint constraint analyses on various cosmological
models by integrating GRBs standardized with the Amati relation
alongside other well-established cosmological probes. Their study
validated the consistency among different observational datasets
and provided evidence supporting the dark energy dynamics and the
existence of spatial curvature
\cite{Cao:2021MNRAS.501.1520C,Cao:2022MNRAS.513.5686C,Cao:2023PhRvD.107j3521C}.

\subsubsection{Cosmology with the Yonetoku relation}

Kodama et al. \cite{Kodama:2008MNRAS.391L...1K} calibrated the
Yonetoku relation in a model-independent way with 33 LGRBs with
known redshifts below 1.62. With additional 30 GRBs with redshifts
ranging from 1.8 to 5.6, they applied the calibrated Yonetoku
relation on a flat universe to get $\Omega_{\rm
m}=0.37^{+0.14}_{-0.11}$ at the 1$\sigma$ confidence level. For
the non-flat $\Lambda$CDM model, the most likely parameters are
derived as $\Omega_{\rm m}=0.25^{+0.27}_{-0.14}$ and
$\Omega_{\Lambda}=1.25^{+0.10}_{-1.25}$. Using a similar
calibration method, Tsutsui et al.
\cite{Tsutsui:2009JCAP...08..015T} compared the cosmology
parameters derived by using the Yonetoku relation and the Amati
relation. They found that the results of the two methods are
consistent at the 2$\sigma$ level, but are obviously different at
the 1$\sigma$ level. Additionally, the Hubble diagrams derived
from these two relations exhibit a systematic difference in the
high-redshift region. In another study, Tsutsui et al.
\cite{Tsutsui:2009MNRAS.394L..31T} also used 63 GRBs and 192 SNe
to investigate the flat $\omega$CDM and CPL models. They found
that the resultant cosmology parameters are consistent with the
standard $\Lambda$CDM model at the 2$\sigma$ level.

\subsubsection{Cosmology with the Dainotti relation}

The usage of the Dainotti relation as an independent cosmology
probe was first proposed by Cardone et al
\cite{Cardone:2010MNRAS.408.1181C}. They gathered a sample of 66
GRBs with a plateau phase in their X-ray afterglow and constructed
the Hubble diagram with the data set. The standard $\Lambda$CDM
model, $\omega$CDM model and CPL model were tested by them. For
the $\Lambda$CDM model and $\omega$CDM model, the results derived
by using the Dainotti relation are consistent with previous
studies, indicating that the GRB data does not introduce
systematic biases on the cosmology parameters and affirming the
effectiveness of the afterglow relation in cosmology researches.
For the CPL model, the cosmology parameters are not well
constrained, but are still in broad consistency with the standard
$\Lambda$CDM model at the 1$\sigma$ confidence level. Using a
sample of 68 LGRBs, Postnikov et al.
\cite{Postnikov:2014ApJ...783..126P} applied the Dainotti relation
to study the evolution of the dark energy EOS parameter $\omega$
with redshift. They adopted a non-parametric method. The
observational data of SNe Ia and baryonic acoustic oscillations
(BAOs) are also incorporated. The $\omega$ parameter derived by
them is close to the constant of $-1$, consistent with the
standard $\Lambda$CDM model.

Wang et al. \cite{Wang:2022ApJ...926..178W} analyzed 31 LGRBs with
a plateau phase in the X-ray light curve. For all these events,
the post-plateau decay follows a temporal power-law index of $-2$.
Such a decaying pattern can be well described by the energy
injection process due to the spin-down of a magnetar. Since the
same energy injection mechanism is likely involved in all these
GRBs, they argued that these GRBs could serve as more reliable
standard candles. After standardizing the sample using the
Dainotti relation, they applied these GRBs to study various
cosmology models. The Hubble diagram derived by them are generally
consistent with previous results across both low and high
redshifts. Interestingly, a $3\sigma$-level evidence for cosmic
acceleration was revealed. Hu et al. \cite{Hu:2021MNRAS.507..730H}
utilized the OHD to alleviate the circularity
problem and constrained the cosmology parameters by combining 31
LGRBs from Wang et al. \cite{Wang:2022ApJ...926..178W}. 5
additional SGRBs, whose X-ray afterglow is dominated by magnetic
dipole radiation, are also included. For the non-flat $\Lambda$CDM
model, the best-fit parameters are $\Omega_{\rm m} =
0.33^{+0.06}_{-0.09}$ and $\Omega_{\Lambda}=1.06^{+0.15}_{-0.34}$
at the 1$\sigma$ level. Further considering the Pantheon SNe Ia
sample, they obtained $\omega=-1.11^{+0.11}_{-0.15}$ and
$\Omega_{\rm m}=0.34^{+0.05}_{-0.04}$ for the flat $\omega$CDM
universe. They concluded that the standard $\Lambda$CDM model
still remains to be the preferred model. Additionally, the
Dainotti relation drawn from optical and radio samples has also
been used to investigate the universe
\cite{Tian:2023ApJ...958...74T,Li:2024A&A...689A.165L}, which
gives consistent results with that using X-ray afterglows.

Cao et al. \cite{Cao:2022MNRAS.510.2928C} utilized three
datasets from Wang et al. \cite{Wang:2022ApJ...926..178W} and Hu
et al. \cite{Hu:2021MNRAS.507..730H} to apply the 2D Dainotti
relation to constrain the six cosmological models. Despite the
relatively weak constraints derived from the 2D Dainotti relation,
they remain consistent with those obtained from the OHD $+$ BAO
data, yielding $\Omega_m \sim 0.3$, thus demonstrating the
potential of the Dainotti relation for standardizing GRBs. Using
the platinum sample defined by Dainotti et al.
\cite{Dainotti:2016ApJ...825L..20D,Dainotti:2017ApJ...848...88D},
Cao et al. \cite{Cao:2022MNRAS.512..439C} confirmed the
reliability of the 3D Dainotti relation in standardizing GRBs. The
combination of the 3D Dainotti relation and the Amati relation
yields better constraints compared to the 3D Dainotti relation
alone. However, the precision remains inferior to that achieved by
the OHD $+$ BAO data, indicating that current GRB data alone are
insufficient for providing strict cosmological constraints
\cite{Cao:2022MNRAS.512..439C}. Moreover, a detailed comparative
analysis indicated that the 3D Dainotti relation outperforms the
2D Dainotti in standardizing GRBs, exhibiting improved parameter
consistency across various cosmological models
\cite{Cao:2022MNRAS.516.1386C}.

\subsubsection{Cosmology with the $L_{\rm X}-T_{\rm a}-E_{\rm p}$ correlation}

\begin{figure}[H]
\includegraphics[width=7.2 cm]{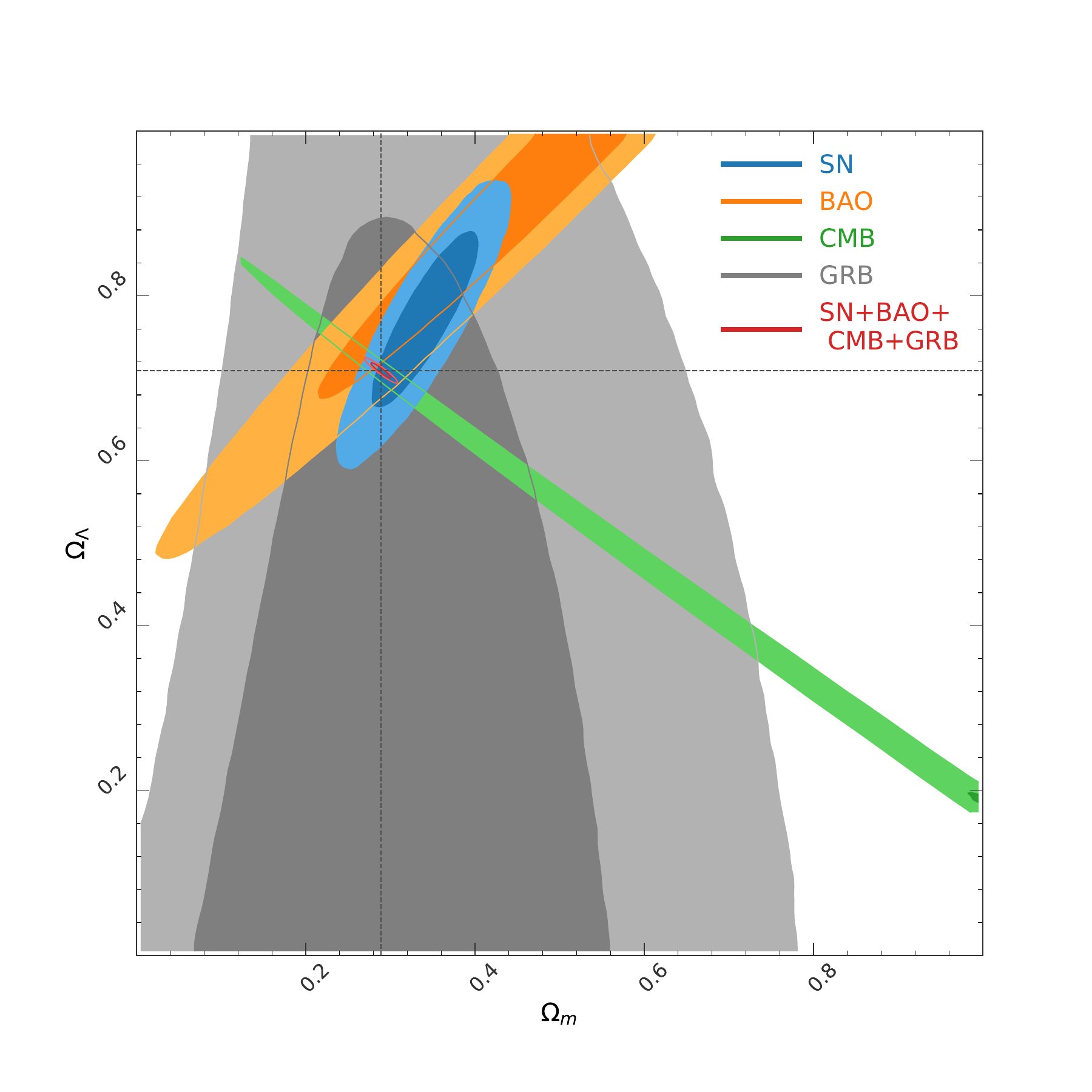}
\includegraphics[width=7.2 cm]{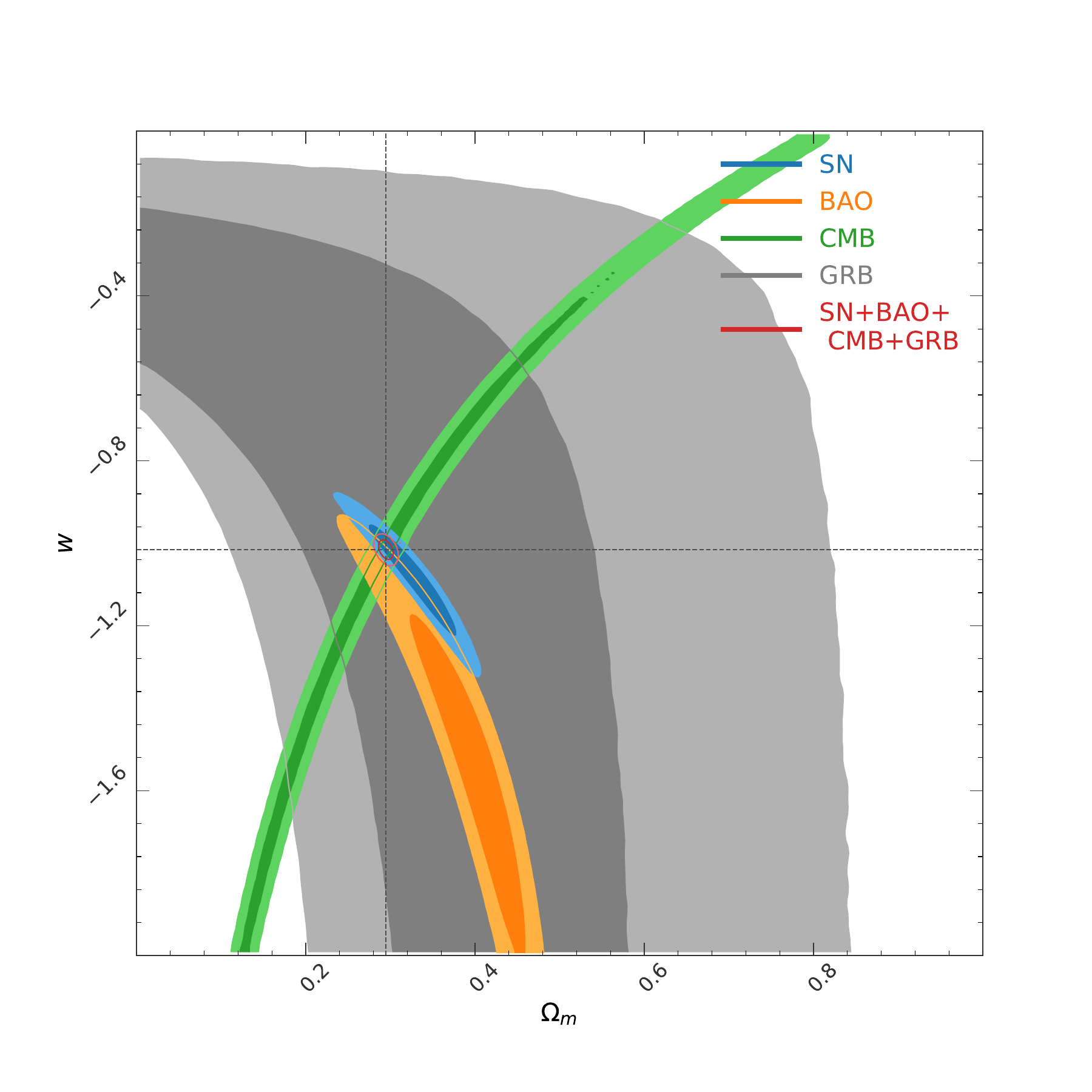}
\caption{The joint constraints for the non-flat $\Lambda$CDM model
(left-hand panel) and flat $\omega$CDM model (right-hand panel) by
Xu et al. \cite{Xu:2021ApJ...920..135X}. The observational data of
SNe Ia, BAO, CMB, and the $L_{\rm X}-T_{\rm a}-E_{\rm p}$ correlation
of GRBs are depicted in blue, orange, green, and gray, respectively.
The dashed lines indicate the best-fit values for the two models:
$\Omega_{\rm m}=0.289\pm0.008$ and $\Omega_{\Lambda}=0.710\pm0.006$
for the non-flat $\Lambda$CDM model, and $\Omega_{\rm m}=0.295\pm0.006$
and $\omega=-1.015\pm0.015$ for the flat $\omega$CDM model.\label{fig2}}
\end{figure}

Xu et al. \cite{Xu:2021ApJ...920..135X} argued that the $L_{\rm
X}-T_{\rm a}-E_{\rm iso}$ correlation is redshift-dependent.
Furthermore, since both $L_{\rm X}$ and $E_{\rm iso}$ depends on
the redshift in a similar form, this correlation is insensitive to
cosmology parameters and thus is unsuitable for cosmology study.
They then combined the Amati relation and the $L_{\rm X}-T_{\rm
a}-E_{\rm iso}$ correlation to derive a new three-parameter
correlation as $L_{\rm X} \propto T_{\rm a}^{-1.08 \pm 0.8} E_{\rm
p}^{0.76 \pm 0.14}$, with an intrinsic scatter of $\sigma_{\rm
int}=0.54 \pm 0.04$. This expression is named as the $L_{\rm
X}-T_{\rm a}-E_{\rm p}$ correlation, which is also affected by
redshift evolution. Xu et al. \cite{Xu:2021ApJ...920..135X} used
the EP method to mitigate this effect. The $L_{\rm X}-T_{\rm
a}-E_{\rm p}$ correlation is somewhat analogous to the Combo
relation. Although the $L_{\rm X}-T_{\rm a}-E_{\rm p}$ correlation
itself is insufficient to constrain cosmology parameters
satisfactorily, Xu et al. \cite{Xu:2021ApJ...920..135X} argued
that it can be combined with the cosmic microwave background
(CMB), SNe Ia, and BAO data to constrain the parameters of various
cosmological models, as shown in Figure \ref{fig2}. They derived
$\Omega_{\rm m}=0.291\pm0.005$ for the flat $\Lambda$CDM model,
$\Omega_{\rm m}=0.289\pm0.008$ and $\Omega_{\Lambda}=0.710\pm0.006$
for the non-flat $\Lambda$CDM model, and $\Omega_{\rm m}=0.295\pm0.006$
and $\omega=-1.015\pm0.015$ for the flat $\omega$CDM model.

Interestingly, for optical afterglows with a plateau phase, there
also exists a three parameter correlation similar to the $L_{\rm
X}-T_{\rm a}-E_{\rm p}$ relation. The optical afterglow sample
have also been utilized to investigate the $X_{1}X_{2}$CDM model,
in which the dark energy is comprised of two independent
components \cite{Li:2024A&A...689A.165L}. According to the
Bayesian information criterion, the $\omega$CDM model with a
single dark energy component is preferred over the $X_{1}X_{2}$CDM
model.

\subsubsection{Cosmology with the Combo relation}

Izzo et al. \cite{Izzo:2015A&A...582A.115I} employed the Combo
relation to constrain the parameters of various cosmology models,
using a calibration technique designed to minimize the dependence
on SNe Ia. Adopting 60 LGRBs alone, they obtained the best-fit
parameters as $\Omega_{\rm m}=0.29^{+0.23}_{-0.15}$ for the flat
$\Lambda$CDM model, and $\Omega_{\rm m}=0.40^{+0.22}_{-0.16}$,
$\omega=-1.43^{+0.78}_{-0.66}$ for the flat $\omega$CDM model.
Here the error ranges are all corresponding to the 1$\sigma$
level. Note that their results for the CPL model align with the
standard $\Lambda$CDM universe, although the uncertainties of the
derived parameters are still relatively large. Muccino et al.
\cite{Muccino:2021ApJ...908..181M} expanded the sample size to 174
LGRBs. They utilized the Combo relation to obtain a much tighter
constraints on the cosmological parameters. When using $H_0=(74.03
\pm 0.5)\ \rm{km\ s^{-1}\ Mpc^{-1}}$
\cite{Riess:2019ApJ...876...85R} as a prior for the Hubble
constant, the updated parameters are $\Omega_{\rm m} =
0.32^{+0.05}_{-0.05}$ for a flat $\Lambda$CDM universe, and
$\Omega_{\rm m} = 0.34^{+0.08}_{-0.07}$, $\Omega_{\Lambda} =
0.91^{+0.22}_{-0.35}$ for a non-flat $\Lambda$CDM model. On the
other hand, when they use $H_0=(67.4 \pm 1.42)\ \rm{km\ s^{-1}\
Mpc^{-1}}$ \cite{PlanckCollaboration:2020A&A...641A...6P} as the
prior, the derived parameters are $\Omega_{\rm m} =
0.22^{+0.04}_{-0.03}$ for the flat $\Lambda$CDM universe, while
$\Omega_{\rm m} = 0.24^{+0.06}_{-0.05}$, $\Omega_{\Lambda} =
1.01^{+0.15}_{-0.25}$ for the non-flat $\Lambda$CDM universe.
Moreover, Muccino et al. \cite{Muccino:2021ApJ...908..181M}
demonstrated that the $\omega(z)$ parameter is consistent with the
$\Lambda$CDM model in the low-redshift region. However, at high
redshifts, $\omega(z)$ deviates from the standard value of $-1$.

Khadka et al. \cite{Khadka:2021JCAP...09..042K} conducted
a comprehensive analysis of five datasets. They used the Combo
relation to constrain the six cosmological models, employing the
simultaneous fitting technique to bypass the circularity problem.
The results revealed moderate inconsistencies in the parameters of
the Combo relation across various cosmological models, leading to
the conclusion that the five Combo relation datasets analyzed in
the study are unreliable for constraining cosmological parameters.
Furthermore, the intrinsic dispersion parameter of the dataset is
found unsuitable to act as a metric for evaluating the robustness
of cosmological constraints \cite{Khadka:2021JCAP...09..042K}.

\subsection{Pseudo-redshifts calculated by using the correlations}

After over 20 years of observations, redshift information has been
obtained for a substantial number of GRBs, allowing us to infer
their luminosity distances and to utilize them for cosmological
researches. However, for many GRBs, the redshifts are still
unavailable. The most distant GRB has a redshift of approximately
9.4 \cite{Cucchiara:2011ApJ...736....7C}, yet some GRBs without
redshift measurements may originate from even earlier cosmic
epochs. Empirical correlations observed in the prompt and
afterglow phases may offer insights on the redshifts of these
GRBs. However, most of these correlations exhibit considerable
dispersion, leading to significant uncertainty in the derived
redshift. So, the redshifts calculated by using these empirical
correlations are called pesudo-redshifts. In this subsection, we
briefly summarize various efforts trying to calculate
pseudo-redshifts from GRB correlations.

\subsubsection{A redshift indicator related to the Amati relation}

Based on the Amati relation, Atteia
\cite{Atteia:2003A&A...407L...1A} found that a combination of some
observed quantities in GRB prompt phase shows a weak dependence on
the isotropic equivalent energy, suggesting its potential as a
redshift indicator. The combined parameter, denoted as $X$, is
defined as $X=n_{\gamma}/(E_{\rm p}T_{90}^{0.5})$, where
$n_\gamma$ represents the observed photon numbers between $E_{\rm
p}$/100 and $E_{\rm p}$/2. $X$ depends on the redshift as
\begin{eqnarray}
X = A \times f(z),
\end{eqnarray}
where $A$ is a normalization parameter and $f(z)$ refers to the
redshift evolution of $X$ derived from a ``standard'' GRB in a
standard $\Lambda$CDM universe. The energy spectrum of prompt GRB
emission can be described by the so called Band function, with
typical power-law indices of $\alpha=-1.0$, $\beta=-2.3$, and
characteristic energy of $E_{\rm p}=250$ keV. The pseudo-redshift,
$z_{\rm est}$, can then be calculated as $z_{\rm est} =
f^{-1}(X)/A$. The pseudo-redshift calculated in this way may have
an uncertainty up to a factor two, but it allows one to quickly
identify high-redshift GRBs.

\subsubsection{Pseudo-redshifts from the Yonetoku relation}

The pseudo-redshift of a GRB can also be derived by inversely
applying the best-fitting Yonetoku relation. For this purpose,
Yonetoku et al. \cite{Yonetoku:2004ApJ...609..935Y} set a flux
threshold to ensure a higher signal-to-noise ratio, excluding
those weak GRBs whose pseudo-redshift could not be credibly
determined. A sample of 689 BATSE GRBs, with time-averaged energy
spectra in the prompt phase described by the Band function
\cite{Band:1993ApJ...413..281B} were collected. The
pseudo-redshifts were calculated for them, and were further used
to investigate the evolution of the GRB formation rate over
redshift.

Additionally, Tsutsui et al. \cite{Tsutsui:2013MNRAS.431.1398T}
derived the best-fit Yonetoku relation from 8 well-defined SGRBs.
Their results can be expressed as
\begin{eqnarray}\label{func17}
\frac{d_L^2}{(1+z)^{1.59}} = \frac{10^{52.29} \, \text{erg s}^{-1}}{4 \pi F_p}
\left( \frac{E_{\rm p,\text{obs}}}{774.5 \, \text{keV}} \right)^{1.59},
\end{eqnarray}
where $F_p$, and $E_{\rm p,obs}$ are the peak flux and peak energy
of the prompt energy spectrum in the observer's frame, respectively.
The right hand side of Equation (\ref{func17}) involves observational
quantities in the prompt phase, while the left hand side is a
function of redshift. Using this equation, the pseudo-redshifts can
be easily determined. They applied the method to a sample of 71 bright
BATSE SGRBs and found that their pseudo-redshifts range from 0.097 to
2.258, with a mean of 1.05.

\subsubsection{Pseudo-redshift from the Dainotti relation}

Dainotti et al. \cite{Dainotti:2011ApJ...730..135D} explored the
potential of using the Dainotti relation, involving only afterglow
parameters, to calculate the pseudo-redshift. For this purpose,
the best-fit Dainotti relation was re-organized and expressed as
\begin{equation}\label{func18}
\begin{aligned}
\log L_{\rm X} &= \log(4 \pi F_{X}) + 2 \log d_L -
(1 - \beta_{\rm a}) \log (1 + z) \\
&= a \log \left( \frac{T_{\rm a}}{1 + z} \right) + b,
\end{aligned}
\end{equation}

where $a$ and $b$ are the coefficient parameters in the Dainotti
relation, $\beta_{\rm a}$
denotes the spectral index of the X-ray afterglow, and
$F_{\rm X}$ is the X-ray flux at the end of the plateau phase. By
solving the above nonlinear function for a particular GRB, the
pseudo-redshifts can be obtained. However, Dainotti et al.
\cite{Dainotti:2011ApJ...730..135D} demonstrated that the
pseudo-redshifts derived from Equation
(\ref{func18}) are somewhat uncertain due to significant
influences from the errors in the observed quantities, the
precision of fitting parameters $a$ and $b$, and the intrinsic
dispersion $\sigma_{\rm int}$, which contributes nonlinearly to
the overall error. These issues may be alleviated with a larger,
high-quality GRB sample. Moreover, the 3D Dainotti relation
\cite{Dainotti:2016ApJ...825L..20D,Dainotti:2017ApJ...848...88D},
combined with the platinum sample selected through strict
criteria, may help give a much better estimate of the redshift due
to its significantly smaller intrinsic dispersion as compared to
the 2D Dainotti relation.

\subsubsection{Pseudo-redshifts from the $L_{\rm X}-T_{\rm a}-E_{\rm iso}$ correlation}

Deng et al. \cite{Deng:2023ApJ...943..126D} argued that the three
parameter $L_{\rm X}-T_{\rm a}-E_{\rm iso}$ correlation can also
be used to calculate the pseudo-redshift. They collected a sample
of 210 GRBs with a plateau phase in the X-ray afterglow, all with
known redshift. They referred the sample as the Starting Sample,
from which a best-fit $L_{\rm X}-T_{\rm a}-E_{\rm iso}$
correlation is updated by using the MCMC method. A redshift
correlation function, g(z), is then derived based on the $L_{\rm
X}-T_{\rm a}-E_{\rm iso}$ correlation as
\begin{equation}\label{func19}
\begin{aligned}
g(z) &= \log \left( \frac{L_X}{10^{47} \, \text{erg s}^{-1}} \right) - a - b \log
\left( \frac{T_a}{10^3 \, \text{s}} \right) - c \log \left( \frac{E_{\text{iso}}}{10^{53}\,
\text{erg}} \right) \\
&= (2 - 2c) \log \left( \int_0^z \frac{1}{\sqrt{(1 + z')^3 \Omega_m + \Omega_\Lambda}} dz' \right)
+ (b + c + \beta_X - c \alpha_\gamma) \log (1 + z) \\
&\quad - c \log (4 \pi S) + \log (4 \pi F_{X}) - a - b \log (T_{a,\text{obs}}) \\
&\quad + 3b - 47 + 53c + (2 - 2c) \log \left( \frac{C}{H_0}
\right),
\end{aligned}
\end{equation}
where $T_{\rm a,obs}$ is the end time of the plateau in the
observer's frame, and $a$, $b$, and $c$ are coefficient parameters
in the $L_{\rm X}-T_{\rm a}-E_{\rm iso}$ correlation.
$S$ and $\alpha_{\gamma}$ represent
the fluence in 15 -- 150 keV and the photon spectral index as
observed by \textit{Swift} BAT, respectively. Deng et al.
\cite{Deng:2023ApJ...943..126D} then collected another sample of
130 GRBs with a plateau phase but without redshift measurement,
which is called the Target Sample. They tried to calculate the
pseudo-redshifts of these Target Sample GRBs by using the $L_{\rm
X}-T_{\rm a}-E_{\rm iso}$ correlation. Substituting the observed
$F_{X}$, $S$, $T_{a,\text{obs}}$ and $\alpha_{\gamma}$ values of
the Target sample into Equation (\ref{func19}), the redshifts can
be conveniently derived. In the process, it was found that a
reasonable solution could not be obtained for 22 GRBs, which means
the pseudo-redshift is unavailable in context of the $L_{\rm
X}-T_{\rm a}-E_{\rm iso}$ correlation. Ultimately, the
pseudo-redshifts are calculated for 108 GRBs by using this method.
Deng et al. \cite{Deng:2023ApJ...943..126D} went further to
utilize both the Starting Sample and the Target Sample to check
the Dainotti relation and the $\overline{L}_{\gamma,90}-L_{\rm X}$
correlation, where $\overline{L}_{\gamma,90}$ = $E_{\rm
iso}/T_{90}$ represents the average luminosity in the prompt
phase. The best-fit Dainotti relations obtained from the samples
are consistent with each other within $1\sigma$ error. However,
although a tight $\overline{L}_{\gamma,90}-L_{\rm X}$ correlation
is observed in the Target Sample, its slope differs significantly
from that of the Starting Sample. The cause of this difference is
still unknown and warrants further investigation in the future.

\section{Conclusions}
\label{sec4}

In this review, several empirical correlations in GRBs, covering
various observed parameters involving both prompt and afterglow
phases, are introduced. The correlations reveal key relationships
among GRB parameters, providing constraints on the radiation
mechanisms and insights into the composition and structure of GRB
jets. Possible physical interpretations of these correlations are
discussed. Their applications in cosmology researches are also
highlighted. Finally, various approaches of calculating the
pseudo-redshifts of GRBs based on the observed correlations are
briefly outlined.

It is interesting to note that for some empirical correlations,
the detailed expressions are different for long GRBs and short
GRBs, indicating inherent population differences possibly related
to their central engines or progenitors. On the other hand, for
several other correlations, both long GRBs and short GRBs have the
same function form, highlighting their shared features. Despite
these insights, it remains challenging to use empirical
correlations as definitive model discriminators due to the
uncertain physical interpretations of these relationships.
Therefore, further theoretical investigations, including detailed
studies of GRB dynamics, radiation processes are essential.
Numerical simulations of the jet dynamics and the particle
acceleration will also be helpful. Additionally, the afterglow
data could provide important constraints on the GRB radiation
efficiency and jet structure. Note that many afterglow
correlations pertain to the plateau phase, which strongly
indicates the plateau phase may have a unique origin. Precise
multi-wavelength afterglow observations beyond the X-ray data
could yield new insights. For example, breaks in the afterglow
light curves, either achromatic or chromatic, could provide key
information on the beaming effects and/or the radiation
mechanisms.

GRBs could potentially help probe the universe at high-redshift,
enabling access to earlier cosmic epochs than SNe Ia. However,
employing GRBs as standard candles requires reliable empirical
correlations, whose robustness depends closely on the physical
classification of GRBs. Additionally, some physical parameters
involved in the correlations may evolve with redshift or be
affected by selection effects related to detector sensitivity
\cite{Dainotti:2018PASP..130e1001D}. Therefore, effective
calibrations are critical to correct for such effects and to
effectively constrain cosmological parameters.

A widely used method for standardizing GRBs involves
calibrating them with SNe Ia data at low redshifts
\cite{Kodama:2008MNRAS.391L...1K,Demianski:2017A&A...598A.112D,
Xu:2021ApJ...920..135X,Tsutsui:2009JCAP...08..015T}, partially
mitigating the circularity problem. However, the joint analysis of
SNe Ia and GRBs assigns significant statistical weights to SNe Ia,
leading to results dominated by SNe Ia and compromising the
independence of GRBs as a cosmological probe. Currently, some
empirical correlations, such as the Amati and Dainotti relations,
are recognized as cosmology-independent and serve as reliable
tools for standardizing GRBs
\cite{Khadka:2020MNRAS.499..391K,Khadka:2021JCAP...09..042K,Cao:2022MNRAS.510.2928C,
Cao:2022MNRAS.512..439C,Cao:2022MNRAS.516.1386C}. Nonetheless,
even high-quality datasets, such as A118 sample
\cite{Khadka:2020MNRAS.499..391K,Khadka:2021JCAP...09..042K,
Liu:2022ApJ...931...50L,FanaDirirsa:2019ApJ...887...13F,Cao:2024JCAP...10..093C,Cao:2023PhRvD.107j3521C},
yield relatively weak cosmological constraints. They are
consistent with the constraints from OHD $+$ BAO data but fail to
substantially enhance the precision. While GRB data provide
valuable complementary constraints to other well-established
cosmological probes, improving the precision requires
higher-quality data. Striking a balance between intrinsic
dispersion and sample size is essential
\cite{Cao:2022MNRAS.516.1386C}.

Employing empirical correlations as redshift indicators for GRBs
presents an intriguing approach. Nevertheless, limitations in
sample quality, sample size, and intrinsic scatter of the
correlations make them more suitable for statistical analyse of
large samples rather than for reliably estimating the redshift of
individual GRBs. Rigorous sample selection and exploration of
novel fitting methods could enhance the reliability of empirical
correlations for this purpose. Furthermore, machine learning
approaches for GRB redshift estimation offer a promising direction
for future researches
\cite{Dainotti:2024ApJ...967L..30D,Bargiacchi:2024arXiv240810707B,
Aldowma:2024MNRAS.529.2676A,Narendra:2024arXiv241013985N}.

The recently launched Chinese-French satellite, Space-based
multiband astronomical Variable Objects Monitor (SVOM)
\cite{Wei:2016arXiv161006892W}, and the Einstein Probe
\cite{Yuan:2015arXiv150607735Y}, along with the upcoming Transient
High-Energy Sky and Early Universe Surveyor (THESEUS) space
missions \cite{Amati:2018AdSpR..62..191A}, will discover more GRBs
and will help us build high-quality GRB samples. The robustness of
currently known empirical correlations is expected to be enhanced,
and more new correlations may even be revealed, which would
facilitate more precise investigation of the cosmic evolution
history at high redshifts \cite{Wang:2015NewAR..67....1W}, address
the Hubble tension \cite{Hu:2023Univ....9...94H}, and refine GRB
classification \cite{Zhang:2006Natur.444.1010Z}.

\acknowledgments{The authors are grateful to the anonymous
referees for helpful suggestions. C. D. thanks J. P. Hu for
valuable discussions. This study was supported by the National
Key R\&D Program of China (2021YFA0718500), the National Natural
Science Foundation of China (Grant Nos. 12233002 and 12041306),
the National SKA Program of China No. 2020SKA0120300, and the
Natural Science Foundation of Xinjiang Uygur Autonomous Region
(No. 2022D01A363). YFH also acknowledges the support from the
Xinjiang Tianchi Program. AK also acknowledges the support from
the Tianchi Talents Project of Xinjiang Uygur Autonomous Region.}

\conflictsofinterest{The authors declare no conflicts of interest.}
\begin{adjustwidth}{-\extralength}{0cm}

\reftitle{References}


\bibliography{ref}

\PublishersNote{}
\end{adjustwidth}
\end{document}